\documentclass[manuscript,numberedappendix,appendixfloats]{emulateapj}

\newcommand{\as}{$^{\prime\prime}$}
\newcommand{\ci}{[C{\sc{I}}]}

\newcommand{\nii}{[N{\sc{II}}]}
\newcommand{\ls}{L$_{\odot}$}
\newcommand{\ms}{M$_{\odot}$}
\newcommand{\kms}{$\,\rm km\,s^{-1}$}

\newcommand{\eqq}{\!=\!}

\newcommand{\jone}{{$J\eqq$ 1$-$0}} 
\newcommand{\jtwo}{{$J\eqq$ 2$-$1}}
\newcommand{\jthree}{{$J\eqq$ 3$-$2}}
\newcommand{\jfour}{{$J\eqq$ 4$-$3}}
\newcommand{\jfive}{{$J\eqq$ 5$-$4}}
\newcommand{\jsix}{{$J\eqq$ 6$-$5}}
\newcommand{\jseven}{{$J\eqq$ 7$-$6}}
\newcommand{\jeight}{{$J\eqq$ 8$-$7}}
\newcommand{\jnine}{{$J\eqq$ 9$-$8}}
\newcommand{\jten}{{$J\eqq$ 10$-$9}}
\newcommand{\jeleven}{{$J\eqq$ 11$-$10}}
\newcommand{\jtwelve}{{$J\eqq$ 12$-$11}}
\newcommand{\jthirteen}{{$J\eqq$ 13$-$12}}

\newcommand{\lfir}{$L_{\rm FIR}$}
\newcommand{\lir}{$L_{\rm IR}$}
\newcommand{\lprimeco}{$L^\prime_{\rm CO}$}
\newcommand{\lco}{$L_{\rm CO}$}

\slugcomment{}

\shorttitle{\lprimeco/\lfir\ Relations Across the {\it Herschel} SPIRE Archive}
\shortauthors{Kamenetzky et al.}

\usepackage{color}
\usepackage{natbib}
\usepackage{bm}
\usepackage{hyperref}
\usepackage{listings}
\begin{document}

\title{\lprimeco/\lfir\ Relations with CO Rotational Ladders of Galaxies Across the {\it Herschel} SPIRE Archive}

\author{J. Kamenetzky\altaffilmark{1}, N. Rangwala\altaffilmark{2}, J. Glenn\altaffilmark{3}, P. R. Maloney\altaffilmark{3}, A. Conley\altaffilmark{3}}
\altaffiltext{1}{Steward Observatory, University of Arizona, 933 North Cherry Avenue, Tucson, AZ 85721}
\altaffiltext{2}{NASA Postdoctoral Fellow, NASA Ames Research Center/ Oak Ridge Associated Universities, Naval Air Station, Moffett Field, Mountain View, CA 94035}
\altaffiltext{3}{University of Colorado at Boulder, Center for Astrophysics and Space Astronomy, 389-UCB, Boulder, CO, USA}
\email{jkamenetzky@as.arizona.edu}

\begin{abstract}
We present a catalog of all CO (\jfour\ through \jthirteen), \ci, and \nii\ lines available from extragalactic spectra from the {\it Herschel} SPIRE Fourier Transform Spectrometer (FTS) archive combined with 
observations of the low-$J$ CO lines from the literature and from the Arizona Radio Observatory. 
This work examines the relationships between \lfir, \lprimeco, and \lco/$L_{\rm CO,1-0}$. We also present a new method for 
estimating probability distribution functions (PDFs) from marginal signal-to-noise ratio {\it Herschel} FTS spectra, which takes into account the instrumental ``ringing" and the resulting highly correlated nature of the spectra. 
The slopes of log(\lfir) vs. log(\lprimeco) are linear for all mid- to high-$J$ CO lines and slightly sublinear if restricted to (U)LIRGs.
The mid- to high-$J$ CO luminosity relative to CO \jone\ increases with increasing \lfir, indicating higher excitement of the molecular gas, though these ratios do not exceed $\sim$ 180.
For a given bin in \lfir, the luminosities relative to CO \jone\ remain relatively flat from \jsix\ through \jthirteen, across three orders of magnitude of \lfir. 
A single component theoretical photon-dominated region (PDR) model cannot match these flat SLED shapes, though combinations of PDR models with mechanical heating added qualitatively match the shapes,
indicating the need for further comprehensive modeling of the excitation processes of warm molecular gas in nearby galaxies.
\end{abstract}


\section{Introduction}\label{sec:intro}

Within the multi-phase interstellar medium (ISM), molecular gas is the most intimately tied to star formation, and therefore to the stellar lifecycle's dramatic effects on galaxy evolution. Though molecular hydrogen is the dominant component of such gas, pure H$_2$ rotational lines are difficult to detect and not particularly sensitive to the low temperatures of most molecular gas. Instead, $^{12}$CO (henceforth CO) and its isotopologues are used to trace the mass, kinematics, and excitation of molecular gas. The ground-level CO \jone\ line is widely used to estimate the total molecular mass in the interstellar medium, and ratios with higher lines provide information on the temperature and density of the emitting gas. Higher lines, however, are increasingly blocked by water absorption in Earth's atmosphere. It was not until the launch of the {\it Herschel} Space Observatory \citep{Pilbratt2010} that the CO ladder up to \jthirteen\ was generally available for the ISM within our Galaxy and in nearby galaxies.

Early SPIRE observations showed much brighter high-$J$ CO emission than would be predicted by cool (T$_{\rm kin} <$ 50 K) molecular gas in giant molecular clouds, the type of gas responsible for the CO \jone\ and other low-$J$ emission \citep[e.g.,][]{Panuzzo2010,Kamenetzky2012,Rangwala2011,Spinoglio2012,Rigopoulou2013,Pereira-Santaella2013}. A warmer, denser (higher pressure) component of molecular gas is responsible for the emission of mid- (\jfour\ to \jsix) to high-$J$ (\jseven\ and above) CO lines (and even warmer emission can be seen in much higher-$J$ lines visible with PACS, as in \citet{Hailey-Dunsheath2012}). UV heating from young O and B stars creates Photodissociation Regions (PDRs), which can reproduce the excitation and emission of the low-$J$ lines. However, PDR models often cannot explain the bright emission seen in high-$J$ lines, which may require mechanical excitation via shocks, turbulence, winds, and other dynamical processes within galaxies. In addition to illuminating the excitation mechanisms of the gas, CO emission is also studied in the context of Kennicutt-Schmidt \citep[K-S law]{Kennicutt1998}, which relate the gas surface density to the star formation rate (SFR) surface density.

Now that {\it Herschel}'s mission is complete, work is underway to examine the full archival data set. 
\citet{Kamenetzky2014} (henceforth K14) presented a two-component modeling procedure for a sample of galaxies observed with the {\it Herschel} SPIRE Fourier Transform Spectrometer (FTS). In the 17 galaxy systems studied in that paper, the warm molecular gas accounted for about 10\% of the total molecular mass, but 90\% of the CO luminosity. Here we expand this sample and compile  a comprehensive, uniformly calibrated set of CO \jfour\ to \jthirteen, \ci\ (609 and 370 $\mu$m), and \nii\ (205 $\mu$m) line fluxes for the galaxies observed by the {\it Herschel} SPIRE FTS, as well as a similarly matched set of CO \jone\ to \jthree\ lines from the literature and the Arizona Radio Observatory (ARO). A future paper will include a full, two-component likelihood analysis of each galaxy's CO SLED, in order to derive cold and warm gas temperatures, densities, and masses, as in K14.

The observations and processing are described in Section \ref{sec:obs}. 
Section \ref{sec:analysis} presents motivation for fitting the relationships between \lprimeco\ and \lfir, our results broken down by subsamples, and 
comparison to the similar studies of \citet{Lu2014}, \citet{Greve2014} and \citet{Liu2015}. 
Discussion of trends and comparisons to theoretical models are in Sections \ref{sec:disc}.

\section{Observations}\label{sec:obs}

We compiled a list of successful extragalactic SPIRE FTS proposals (301 spectra) and searched the {\it Herschel} Science Archive (HSA) for the available data.
In some cases, programs for higher or unknown redshift galaxies did not result in spectra with measurable CO emission, so those observations (about 74) are not presented. Table \ref{tbl:galaxies} lists the basic galaxy information and observation IDs for all galaxies for which at least one FTS line measurement or upper limit is reported.\footnote{ The spectra for this survey came from the following programs, with the total number of observations presented in Table \ref{tbl:galaxies} in parentheses: 
\texttt{OT1\_nlu\_1} (92), \texttt{KPOT\_pvanderw\_1} (31), \texttt{ OT1\_dfarrah\_1} (28), \texttt{OT1\_jsmith01\_1} (23), \texttt{OT1\_pogle01\_1} (13), \texttt{KPGT\_cwilso01\_1} (12), \texttt{OT1\_lyoung\_1} (8), \texttt{GT1\_lspinogl\_2} (10), \texttt{OT1\_pvanderw\_4} (5), \texttt{
GT2\_vleboute\_3} (2), \texttt{OT2\_vkulkarn\_3} (2), \texttt{KPGT\_rguesten\_1} (1), \texttt{
OT1\_dmarrone\_1} (0), \texttt{
OT1\_rivison\_1} (0), \texttt{
OT2\_drigopou\_3} (0), \texttt{
OT2\_rivison\_2} (0).}

\LongTables
\begin{deluxetable*}{lrrrrrrrr}
\tablecaption{Galaxies and Observations Utilized\label{tbl:galaxies}}
\tablehead{\colhead{Galaxy} & \colhead{RA} & \colhead{Dec} & \colhead{Log(\lfir)} & \colhead{$D_{L}$} & \colhead{$n_{\rm det}$} & \colhead{$n_{\rm ul}$} & \colhead{FTS ObsId} & \colhead{Phot ObsID}\\ \colhead{} & \colhead{J2000} & \colhead{J2000} & \colhead{\ls} & \colhead{Mpc} & \colhead{} & \colhead{} & \colhead{ } & \colhead{}}
\startdata
\object{NGC0023} & 0h09m53.36s & +25d55m26.4s & 10.9 & 68 & 5 & 7 & 1342247622 & 1342234681 \\
\object{NGC34} & 0h11m06.55s & $-$12d06m26.3s & 11.2 & 85 & 12 & 0 & 1342199253 & 1342199383 \\
\object{MCG-02-01-051} & 0h18m50.86s & $-$10d22m37.5s & 11.2 & 120 & 8 & 4 & 1342247617 & 1342234694 \\
\object{IC10-B11-1} & 0h20m27.70s & +59d16m59.4s & 7.5 & 1 & 8 & 5 & 1342246982 & 1342201446 \\
\object{IRAS 00188-0856} & 0h21m26.53s & $-$08d39m27.1s & 12.2 & 591 & 4 & 5 & 1342246259 & 1342234693 \\
\object{ESO350-IG038}\tablenotemark{a} & 0h36m52.46s & $-$33d33m17.4s & 10.8 & 87 & 1 & 8 & 1342246978 & 1342199386 \\
\object{NGC205-copeak} & 0h40m24.10s & +41d41m50.4s & 6.1 & 1 & 1 & 9 & 1342212315 & 1342188661 \\
\object{IRAS 00397-1312} & 0h42m15.53s & $-$12d56m02.8s & 12.6 & 1285 & 0 & 7 & 1342246257 & 1342234696 \\
\object{NGC0232a} & 0h42m45.82s & $-$23d33m41.7s & 11.2 & 95 & 8 & 4 & 1342221707 & 1342234699 \\
\object{NGC253} & 0h47m33.12s & $-$25d17m17.6s & 10.3 & 3 & 12 & 1 & 1342210847 & 1342199387 \\
\object{I Zw 1}\tablenotemark{a} & 0h53m34.94s & +12d41m36.2s & 11.4 & 272 & 2 & 9 & 1342238246 & 1342238252 \\
\object{MCG+12-02-001} & 0h54m03.61s & +73d05m11.8s & 11.2 & 72 & 11 & 2 & 1342213377 & 1342199365 \\
\object{NGC0317B} & 0h57m40.37s & +43d47m32.4s & 11.0 & 80 & 8 & 4 & 1342239358 & 1342238255 \\
\object{IRAS 01003-2238} & 1h02m49.90s & $-$22d21m57.3s & 11.9 & 539 & 1 & 7 & 1342246256 & 1342234707 \\
\object{3C 31} & 1h07m24.96s & +32d24m45.2s &  & 75 & 3 & 5 & 1342239344 & 1342236245 \\
\object{IC1623} & 1h07m47.00s & $-$17d30m25.0s & 11.4 & 86 & 11 & 1 & 1342212314 & 1342199388 \\
\object{MCG-03-04-014}\tablenotemark{a} & 1h10m08.92s & $-$16d51m11.1s & 11.4 & 152 & 6 & 6 & 1342213442 & 1342234709 \\
\object{ESO244-G012}\tablenotemark{a} & 1h18m08.26s & $-$44d27m43.0s & 11.1 & 95 & 5 & 6 & 1342221708 & 1342234726 \\
\object{CGCG436-030}\tablenotemark{a} & 1h20m02.58s & +14d21m42.5s & 11.5 & 138 & 9 & 3 & 1342213443 & 1342237499 \\
\object{ESO353-G020} & 1h34m51.29s & $-$36d08m15.0s & 10.8 & 66 & 7 & 5 & 1342247615 & 1342234721 \\
\object{IRASF01417+1651} & 1h44m30.52s & +17d06m08.9s &  & 120 & 7 & 5 & 1342239343 & 1342237555 \\
\object{NGC0695} & 1h51m14.28s & +22d34m55.2s & 11.4 & 143 & 6 & 5 & 1342224767 & 1342238266 \\
\object{Mrk 1014}\tablenotemark{a} & 1h59m50.21s & +00d23m40.6s & 12.3 & 763 & 1 & 7 & 1342238707 & 1342237540 \\
\object{NGC0828} & 2h10m09.50s & +39d11m24.7s & 11.1 & 80 & 4 & 7 & 1342239357 & 1342239822 \\
\object{NGC0877a} & 2h18m00.12s & +14d32m34.2s &  & 57 & 5 & 7 & 1342239342 & 1342238267 \\
\object{NGC 891-1} & 2h22m33.41s & +42d20m56.9s & 10.2 & 10 & 7 & 5 & 1342213376 & 1342189430 \\
\object{UGC01845} & 2h24m07.89s & +47d58m11.3s & 10.9 & 70 & 5 & 7 & 1342240022 & 1342239799 \\
\object{NGC0958} & 2h30m42.80s & $-$02d56m23.7s & 10.9 & 82 & 3 & 7 & 1342239339 & 1342238277 \\
\object{0235+164} & 2h38m38.93s & +16d36m59.3s &  & 4278 & 0 & 1 & 1342249452 & 1342224149 \\
\object{NGC1068} & 2h42m40.71s & $-$00d00m47.8s & 10.9 & 16 & 13 & 0 & 1342213445 & 1342189425 \\
\object{NGC1056} & 2h42m48.30s & +28d34m27.1s & 9.7 & 24 & 5 & 8 & 1342204024 & 1342226630 \\
\object{UGC02238} & 2h46m17.50s & +13d05m44.9s & 11.1 & 93 & 5 & 5 & 1342239340 & 1342238270 \\
\object{NGC1097} & 2h46m19.00s & $-$30d16m30.0s & 10.4 & 16 & 8 & 5 & 1342239337 & 1342188586 \\
\object{UGC02369} & 2h54m01.81s & +14d58m14.3s & 11.4 & 142 & 8 & 3 & 1342239341 & 1342239831 \\
\object{NGC1222} & 3h08m56.74s & $-$02d57m18.5s & 10.4 & 35 & 8 & 5 & 1342239354 & 1342239262 \\
\object{UGC02608} & 3h15m01.24s & +42d02m09.2s & 11.1 & 104 & 4 & 5 & 1342239356 & 1342239819 \\
\object{NGC1266} & 3h16m00.70s & $-$02d25m38.0s & 10.2 & 31 & 13 & 0 & 1342239353 & 1342189424 \\
\object{IRAS 03158+4227}\tablenotemark{a} & 3h19m12.40s & +42d38m28.0s & 12.4 & 623 & 4 & 4 & 1342224764 & 1342226656 \\
\object{3C 84} & 3h19m48.16s & +41d30m42.1s & 10.8 & 78 & 11 & 2 & 1342249054 & 1342203614 \\
\object{NGC1365-SW} & 3h33m35.90s & $-$36d08m35.0s & 10.8 & 21 & 10 & 2 & 1342204021 & 1342201432 \\
\object{NGC1365-NE} & 3h33m36.60s & $-$36d08m20.0s & 10.8 & 21 & 10 & 3 & 1342204020 & 1342201432 \\
\object{NGC1377} & 3h36m39.10s & $-$20d54m08.0s & 9.7 & 24 & 12 & 1 & 1342239352 & 1342189505 \\
\object{NGC1482} & 3h54m38.90s & $-$20d30m09.0s & 10.5 & 25 & 12 & 1 & 1342248233 & 1342189504 \\
\object{IRAS 03521+0028}\tablenotemark{a} & 3h54m42.19s & +00d37m02.0s & 12.3 & 709 & 2 & 7 & 1342238704 & 1342239850 \\
\object{UGC02982} & 4h12m22.53s & +05d32m50.4s & 10.9 & 77 & 3 & 7 & 1342240021 & 1342239938 \\
\object{ESO420-G013}\tablenotemark{a} & 4h13m49.65s & $-$32d00m24.1s & 10.7 & 49 & 8 & 5 & 1342242590 & 1342227719 \\
\object{NGC1572}\tablenotemark{a} & 4h22m42.81s & $-$40d36m03.2s & 11.0 & 86 & 6 & 5 & 1342242588 & 1342227720 \\
\object{IRAS04271+3849} & 4h30m33.10s & +38d55m48.4s & 10.9 & 86 & 6 & 4 & 1342227786 & 1342229106 \\
\object{NGC1614} & 4h33m59.85s & $-$08d34m44.0s & 11.3 & 68 & 12 & 1 & 1342192831 & 1342203628 \\
\object{UGC03094} & 4h35m33.75s & +19d10m17.5s & 11.1 & 108 & 4 & 6 & 1342227522 & 1342239944 \\
\object{MCG-05-12-006}\tablenotemark{a} & 4h52m04.96s & $-$32d59m25.9s & 10.9 & 78 & 6 & 4 & 1342242589 & 1342229237 \\
\object{IRAS F05189-2524}\tablenotemark{a} & 5h21m01.47s & $-$25d21m45.4s & 11.8 & 185 & 10 & 2 & 1342192833 & 1342203632 \\
\object{IRAS05223+1908} & 5h25m16.65s & +19d10m48.5s &  & 130 & 2 & 0 & 1342228738 & 1342229652 \\
\object{MCG+08-11-002} & 5h40m43.65s & +49d41m41.8s & 11.2 & 86 & 9 & 3 & 1342230414 & 1342229112 \\
\object{NGC1961} & 5h42m04.37s & +69d22m41.9s & 10.7 & 61 & 9 & 4 & 1342228708 & 1342227742 \\
\object{UGC03351} & 5h45m48.00s & +58d42m03.7s & 11.1 & 67 & 6 & 6 & 1342230415 & 1342229115 \\
\object{IRAS05442+1732} & 5h47m11.15s & +17d33m47.2s & 11.0 & 81 & 5 & 5 & 1342230413 & 1342229653 \\
\object{IRAS 06035-7102} & 6h02m54.01s & $-$71d03m10.2s & 12.0 & 353 & 10 & 2 & 1342230420 &  \\
\object{UGC03410a} & 6h14m29.64s & +80d26m59.4s & 10.8 & 61 & 3 & 8 & 1342231072 & 1342229131 \\
\object{NGC2146-NW} & 6h18m36.70s & +78d21m32.0s & 10.8 & 17 & 12 & 0 & 1342219554 & 1342191186 \\
\object{NGC2146-nuc} & 6h18m38.60s & +78d21m24.0s & 10.8 & 17 & 11 & 2 & 1342204025 & 1342191186 \\
\object{NGC2146-SE} & 6h18m40.50s & +78d21m16.0s & 10.8 & 17 & 11 & 2 & 1342219555 & 1342191186 \\
\object{IRAS 06206-6315}\tablenotemark{a} & 6h21m01.21s & $-$63d17m23.5s & 12.0 & 411 & 4 & 5 & 1342231038 & 1342226638 \\
\object{ESO255-IG007}\tablenotemark{a} & 6h27m21.63s & $-$47d10m36.3s &  & 166 & 7 & 4 & 1342231084 & 1342226643 \\
\object{UGC03608} & 6h57m34.42s & +46d24m10.7s & 11.1 & 97 & 6 & 6 & 1342228744 & 1342229649 \\
\object{NGC2342b} & 7h09m12.09s & +20d36m13.1s & 10.8 & 77 & 5 & 6 & 1342228730 & 1342230778 \\
\object{NGC2342a} & 7h09m18.05s & +20d38m10.0s & 10.8 & 77 & 5 & 5 & 1342228729 & 1342230778 \\
\object{NGC2369} & 7h16m37.60s & $-$62d20m35.9s & 10.8 & 43 & 10 & 3 & 1342231083 & 1342229670 \\
\object{NGC2388a} & 7h28m53.43s & +33d49m08.4s & 11.0 & 62 & 7 & 6 & 1342231071 & 1342229477 \\
\object{MCG+02-20-003} & 7h35m43.44s & +11d42m34.8s & 10.8 & 72 & 7 & 6 & 1342228728 & 1342229463 \\
\object{IRAS 07598+6508}\tablenotemark{a} & 8h04m30.45s & +64d59m52.2s & 12.1 & 693 & 0 & 10 & 1342253659 & 1342229642 \\
\object{B2 0827+24} & 8h30m52.09s & +24d10m59.8s &  & 5818 & 0 & 1 & 1342253660 & 1342230773 \\
\object{IRAS 08311-2459}\tablenotemark{a} & 8h33m20.60s & $-$25d09m33.7s & 12.2 & 451 & 9 & 1 & 1342230421 & 1342230796 \\
\object{He2-10} & 8h36m15.18s & $-$26d24m33.9s & 9.6 & 10 & 9 & 3 & 1342245083 & 1342196888 \\
\object{IRAS08355-4944} & 8h37m01.86s & $-$49d54m30.0s &  & 110 & 7 & 5 & 1342231975 & 1342226978 \\
\object{NGC2623} & 8h38m24.08s & +25d45m16.6s & 11.4 & 81 & 12 & 0 & 1342219553 & 1342206174 \\
\object{IRAS 08572+3915} & 9h00m25.39s & +39d03m54.4s & 11.8 & 261 & 2 & 8 & 1342231978 & 1342230749 \\
\object{IRAS09022-3615} & 9h04m12.72s & $-$36d27m01.3s & 12.0 & 262 & 11 & 1 & 1342231063 & 1342230799 \\
\object{NGC2764} & 9h08m17.47s & +21d26m36.0s & 10.0 & 40 & 5 & 7 & 1342231057 & 1342245567 \\
\object{NGC2798} & 9h17m22.90s & +41d59m59.0s & 10.4 & 28 & 12 & 1 & 1342252892 & 1342197287 \\
\object{UGC05101} & 9h35m51.65s & +61d21m11.3s & 11.8 & 176 & 9 & 3 & 1342209278 & 1342204962 \\
\object{NGC2976\_00} & 9h47m07.84s & +67d55m52.3s &  & 4 & 1 & 10 & 1342228706 & 1342192106 \\
\object{M81} & 9h55m33.17s & +69d03m55.0s & 9.2 & 4 & 3 & 9 & 1342209851 & 1342185538 \\
\object{M82} & 9h55m52.22s & +69d40m46.9s & 10.4 & 4 & 13 & 0 & 1342208389 & 1342185537 \\
\object{NGC3077} & 10h03m19.10s & +68d44m02.0s & 7.7 & 1 & 1 & 9 & 1342228745 & 1342193015 \\
\object{NGC3110a} & 10h04m02.09s & $-$06d28m28.6s & 11.0 & 73 & 4 & 6 & 1342231971 & 1342234843 \\
\object{3C 236} & 10h06m01.74s & +34d54m10.4s &  & 451 & 0 & 6 & 1342246988 & 1342246613 \\
\object{NGC3221} & 10h22m20.20s & +21d34m22.4s & 10.7 & 61 & 2 & 8 & 1342221714 & 1342246610 \\
\object{NGC3227}\tablenotemark{a} & 10h23m30.58s & +19d51m54.2s & 9.7 & 18 & 12 & 1 & 1342209281 & 1342197318 \\
\object{NGC3256} & 10h27m51.27s & $-$43d54m13.8s & 11.3 & 38 & 13 & 0 & 1342201201 & 1342200126 \\
\object{IRAS 10378+1109} & 10h40m29.17s & +10d53m18.3s & 12.1 & 631 & 2 & 8 & 1342247118 & 1342234867 \\
\object{ESO264-G036} & 10h43m07.68s & $-$46d12m44.9s & 10.9 & 89 & 6 & 4 & 1342249044 & 1342236204 \\
\object{NGC3351} & 10h43m57.70s & +11d42m14.0s & 9.7 & 13 & 9 & 4 & 1342247117 & 1342198885 \\
\object{ESO264-G057} & 10h59m01.82s & $-$43d26m25.9s & 10.7 & 72 & 6 & 5 & 1342249043 & 1342236203 \\
\object{IRASF10565+2448} & 10h59m18.17s & +24d32m34.4s & 11.8 & 192 & 10 & 2 & 1342247096 & 1342234869 \\
\object{NGC3521} & 11h05m48.60s & $-$00d02m09.0s & 10.1 & 12 & 6 & 3 & 1342247743 & 1342198568 \\
\object{IRAS 11095-0238} & 11h12m03.38s & $-$02d54m23.8s & 12.0 & 482 & 4 & 5 & 1342247760 & 1342234863 \\
\object{NGC3627} & 11h20m15.00s & +12d59m30.0s & 10.2 & 12 & 13 & 0 & 1342247604 & 1342198883 \\
\object{NGC3665} & 11h24m43.67s & +38d45m46.0s & 9.7 & 32 & 2 & 7 & 1342247121 & 1342222667 \\
\object{Arp299-B} & 11h28m31.00s & +58d33m41.0s & 11.6 & 49 & 13 & 0 & 1342199249 & 1342199345 \\
\object{Arp299-C} & 11h28m31.00s & +58d33m50.0s & 11.6 & 49 & 13 & 0 & 1342199250 & 1342199345 \\
\object{Arp299-A} & 11h28m33.63s & +58d33m47.0s & 11.6 & 49 & 13 & 0 & 1342199248 & 1342199345 \\
\object{PG 1126-041} & 11h29m16.66s & $-$04d24m07.6s &  & 266 & 1 & 9 & 1342247119 & 1342247271 \\
\object{ESO 320-G030} & 11h53m11.72s & $-$39d07m48.9s & 11.0 & 45 & 12 & 1 & 1342210861 & 1342200129 \\
\object{NGC3982}\tablenotemark{a} & 11h56m28.13s & +55d07m30.9s & 9.8 & 21 & 5 & 5 & 1342209277 & 1342186862 \\
\object{NGC4038} & 12h01m53.00s & $-$18d52m01.0s &  & 23 & 6 & 5 & 1342210860 & 1342188686 \\
\object{NGC4038overlap} & 12h01m54.90s & $-$18d52m46.0s &  & 23 & 8 & 4 & 1342210859 & 1342188686 \\
\object{NGC4051}\tablenotemark{a} & 12h03m09.61s & +44d31m52.8s & 9.5 & 14 & 9 & 4 & 1342209276 & 1342210502 \\
\object{IRAS 12071-0444} & 12h09m45.12s & $-$05d01m13.9s & 12.1 & 591 & 2 & 8 & 1342248239 & 1342234858 \\
\object{NGC4151}\tablenotemark{a} & 12h10m32.58s & +39d24m20.6s &  & 18 & 5 & 7 & 1342209852 & 1342188588 \\
\object{NGC4194} & 12h14m09.63s & +54d31m36.1s & 10.7 & 39 & 10 & 3 & 1342231069 & 1342230869 \\
\object{IRAS12116-5615} & 12h14m22.17s & $-$56d32m32.8s & 11.3 & 115 & 10 & 2 & 1342249462 & 1342226974 \\
\object{NGC4254} & 12h18m49.60s & +14d24m59.0s & 10.8 & 36 & 5 & 4 & 1342236997 & 1342187173 \\
\object{NGC4321} & 12h22m54.90s & +15d49m21.0s & 10.4 & 25 & 8 & 4 & 1342247572 & 1342187322 \\
\object{NGC4388} & 12h25m46.75s & +12d39m43.5s & 10.4 & 38 & 10 & 3 & 1342210849 & 1342248482 \\
\object{NGC4459} & 12h29m00.03s & +13d58m42.8s & 9.1 & 19 & 3 & 7 & 1342248411 & 1342200118 \\
\object{NGC4526} & 12h34m03.03s & +07d41m56.9s & 9.1 & 10 & 6 & 6 & 1342224762 & 1342234889 \\
\object{NGC4536} & 12h34m27.00s & +02d11m17.0s & 10.5 & 27 & 9 & 4 & 1342237025 & 1342189455 \\
\object{NGC4569} & 12h36m49.80s & +13d09m46.0s & 7.5 & 1 & 10 & 3 & 1342248251 & 1342188777 \\
\object{TOL1238-364} & 12h40m52.85s & $-$36d45m21.1s & 10.4 & 46 & 7 & 6 & 1342213381 & 1342202200 \\
\object{NGC4631} & 12h42m08.00s & +32d32m29.0s & 10.4 & 12 & 7 & 3 & 1342247573 & 1342188756 \\
\object{NGC4710} & 12h49m38.96s & +15d09m55.8s & 9.6 & 18 & 5 & 8 & 1342247120 & 1342188766 \\
\object{NGC4736} & 12h50m53.00s & +41d07m14.0s & 9.9 & 8 & 7 & 4 & 1342245851 & 1342188754 \\
\object{Mrk 231} & 12h56m14.23s & +56d52m25.2s & 12.2 & 188 & 11 & 1 & 1342210493 & 1342201218 \\
\object{NGC4826} & 12h56m43.70s & +21d40m58.0s & 9.7 & 9 & 8 & 3 & 1342246992 & 1342188764 \\
\object{MCG-02-33-098} & 13h02m19.80s & $-$15d46m03.5s & 10.7 & 69 & 4 & 7 & 1342247567 & 1342234810 \\
\object{ESO507-G070} & 13h02m52.34s & $-$23d55m17.8s & 11.2 & 91 & 11 & 1 & 1342248421 & 1342234813 \\
\object{NGC5010} & 13h12m26.39s & $-$15d47m51.7s & 10.6 & 43 & 6 & 5 & 1342236996 & 1342234809 \\
\object{IRAS13120-5453} & 13h15m06.35s & $-$55d09m22.7s & 12.0 & 132 & 12 & 0 & 1342212342 & 1342226970 \\
\object{NGC5055} & 13h15m49.30s & +42d01m45.0s & 10.1 & 11 & 5 & 6 & 1342237026 & 1342188753 \\
\object{Arp193} & 13h20m35.34s & +34d08m22.2s & 11.4 & 105 & 12 & 0 & 1342209853 & 1342198191 \\
\object{NGC5104} & 13h21m23.09s & +00d20m33.3s & 10.9 & 82 & 5 & 6 & 1342247566 & 1342236168 \\
\object{MCG-03-34-064}\tablenotemark{a} & 13h22m24.42s & $-$16d43m42.7s &  & 74 & 8 & 3 & 1342249041 & 1342236178 \\
\object{Cen A} & 13h25m27.61s & $-$43d01m08.8s & 9.7 & 4 & 8 & 5 & 1342204037 & 1342188663 \\
\object{NGC5135} & 13h25m44.06s & $-$29d50m01.2s & 11.0 & 58 & 12 & 0 & 1342212344 & 1342202248 \\
\object{ESO 173-G015} & 13h27m23.78s & $-$57d29m22.2s & 11.2 & 39 & 13 & 0 & 1342202268 & 1342203562 \\
\object{NGC5194} & 13h29m52.71s & +47d11m42.6s & 10.4 & 11 & 7 & 6 & 1342201202 & 1342188589 \\
\object{IC4280} & 13h32m53.35s & $-$24d12m25.4s & 10.7 & 70 & 6 & 4 & 1342249042 & 1342236191 \\
\object{M83} & 13h37m00.92s & $-$29d51m56.7s & 10.4 & 7 & 10 & 2 & 1342212345 & 1342188664 \\
\object{Mrk 273} & 13h44m42.11s & +55d53m12.7s & 12.0 & 168 & 11 & 1 & 1342209850 & 1342201217 \\
\object{4C 12.50}\tablenotemark{a} & 13h47m33.36s & +12d17m24.2s & 12.0 & 561 & 1 & 9 & 1342237024 & 1342234792 \\
\object{UGC08739} & 13h49m14.28s & +35d15m19.8s & 10.8 & 76 & 6 & 6 & 1342247123 & 1342236144 \\
\object{ESO221-IG010} & 13h50m56.87s & $-$49d03m18.5s & 10.5 & 39 & 5 & 6 & 1342249461 & 1342238293 \\
\object{Mrk 463}\tablenotemark{a} & 13h56m02.87s & +18d22m19.5s & 11.2 & 226 & 4 & 7 & 1342249047 & 1342236151 \\
\object{M101\_02} & 14h03m41.36s & +54d19m04.9s & 10.1 & 8 & 2 & 10 & 1342230417 & 1342188750 \\
\object{OQ 208}\tablenotemark{a} & 14h07m00.39s & +28d27m14.7s &  & 348 & 3 & 6 & 1342247769 & 1342234785 \\
\object{NGC5653} & 14h30m09.88s & +31d12m56.3s & 10.8 & 55 & 5 & 8 & 1342247565 & 1342236146 \\
\object{IRAS 14348-1447} & 14h37m38.26s & $-$15d00m24.6s & 12.1 & 371 & 8 & 3 & 1342249457 & 1342238301 \\
\object{NGC5713} & 14h40m11.50s & $-$00d17m20.0s & 10.5 & 29 & 8 & 5 & 1342248250 & 1342189520 \\
\object{IRAS 14378-3651} & 14h40m59.01s & $-$37d04m32.0s & 11.9 & 303 & 10 & 1 & 1342227456 & 1342238295 \\
\object{Mrk 478}\tablenotemark{a} & 14h42m07.46s & +35d26m22.9s & 11.1 & 358 & 0 & 9 & 1342238710 & 1342238333 \\
\object{NGC5734a} & 14h45m08.98s & $-$20d52m13.4s & 10.7 & 59 & 3 & 10 & 1342248417 & 1342227731 \\
\object{3C 305}\tablenotemark{a} & 14h49m21.80s & +63d16m15.3s &  & 187 & 1 & 4 & 1342236998 & 1342234915 \\
\object{VV340a}\tablenotemark{a} & 14h57m00.66s & +24d37m05.1s &  & 145 & 6 & 5 & 1342238241 & 1342234779 \\
\object{IC4518ABa} & 14h57m41.15s & $-$43d07m56.2s &  & 68 & 7 & 4 & 1342250514 & 1342239895 \\
\object{NGC5866}\tablenotemark{a} & 15h06m29.50s & +55d45m47.6s & 9.4 & 14 & 7 & 4 & 1342238708 & 1342188749 \\
\object{CGCG049-057} & 15h13m13.09s & +07d13m31.8s & 11.1 & 59 & 11 & 2 & 1342212346 & 1342203077 \\
\object{3C 315} & 15h13m40.08s & +26d07m31.2s &  & 498 & 0 & 4 & 1342239350 & 1342234777 \\
\object{VV705}\tablenotemark{a} & 15h18m06.13s & +42d44m44.5s &  & 181 & 10 & 2 & 1342238712 & 1342229532 \\
\object{ESO099-G004} & 15h24m57.99s & $-$63d07m30.2s & 11.4 & 125 & 10 & 2 & 1342230419 & 1342229209 \\
\object{IRAS 15250+3609} & 15h26m59.40s & +35d58m37.5s & 11.8 & 248 & 2 & 9 & 1342238711 & 1342234775 \\
\object{NGC5936} & 15h30m00.80s & +12d59m21.7s & 10.8 & 61 & 6 & 5 & 1342249046 & 1342238324 \\
\object{Arp220} & 15h34m57.12s & +23d30m11.5s & 12.0 & 81 & 11 & 2 & 1342190674 & 1342188687 \\
\object{NGC5990} & 15h46m16.40s & +02d24m54.7s & 10.7 & 57 & 6 & 7 & 1342240016 & 1342238312 \\
\object{IRAS 15462-0450} & 15h48m56.81s & $-$04d59m33.6s & 12.0 & 456 & 3 & 7 & 1342249045 & 1342238307 \\
\object{3C 326} & 15h52m09.07s & +20d05m48.4s &  & 407 & 0 & 7 & 1342250516 & 1342238327 \\
\object{PKS 1549-79} & 15h56m58.87s & $-$79d14m04.3s &  & 690 & 0 & 8 & 1342253671 & 1342239890 \\
\object{NGC6052} & 16h05m12.94s & +20d32m36.9s & 10.8 & 71 & 6 & 6 & 1342212347 & 1342229560 \\
\object{IRAS 16090-0139} & 16h11m40.48s & $-$01d47m05.6s & 12.3 & 618 & 6 & 4 & 1342238699 & 1342229565 \\
\object{PG 1613+658} & 16h13m57.18s & +65d43m09.6s & 11.5 & 600 & 0 & 8 & 1342242593 & 1342238336 \\
\object{CGCG052-037} & 16h30m56.60s & +04d04m58.3s & 11.1 & 109 & 8 & 2 & 1342251284 & 1342229572 \\
\object{NGC6156} & 16h34m52.50s & $-$60d37m07.7s & 10.8 & 45 & 8 & 4 & 1342231041 & 1342229213 \\
\object{ESO069-IG006} & 16h38m11.84s & $-$68d26m08.5s & 11.7 & 203 & 7 & 4 & 1342231040 & 1342230810 \\
\object{IRASF16399-0937} & 16h42m40.10s & $-$09d43m13.6s & 11.3 & 118 & 8 & 4 & 1342251334 & 1342229188 \\
\object{NGC6240} & 16h52m58.89s & +02d24m03.4s & 11.6 & 108 & 13 & 0 & 1342214831 & 1342203586 \\
\object{IRASF16516-0948} & 16h54m23.81s & $-$09d53m21.4s & 11.0 & 100 & 6 & 5 & 1342251335 & 1342229189 \\
\object{NGC6286b} & 16h58m23.99s & +58d57m20.3s & 11.1 & 85 & 2 & 8 & 1342231068 & 1342229148 \\
\object{NGC6286a} & 16h58m31.56s & +58d56m12.2s & 11.1 & 85 & 7 & 3 & 1342221715 & 1342229148 \\
\object{IRASF17138-1017} & 17h16m35.82s & $-$10d20m41.5s & 11.1 & 76 & 8 & 3 & 1342230418 & 1342229190 \\
\object{IRAS F17207-0014} & 17h23m21.96s & $-$00d17m00.9s & 12.2 & 190 & 11 & 1 & 1342192829 & 1342203587 \\
\object{ESO138-G027} & 17h26m43.30s & $-$59d55m55.6s & 11.1 & 88 & 6 & 6 & 1342231042 & 1342229216 \\
\object{UGC11041} & 17h54m51.82s & +34d46m34.3s & 10.8 & 74 & 4 & 5 & 1342231061 & 1342229169 \\
\object{IRAS17578-0400} & 18h00m31.86s & $-$04d00m53.3s & 11.2 & 62 & 8 & 4 & 1342231047 & 1342229187 \\
\object{NGC6621} & 18h12m55.31s & +68d21m46.8s & 11.0 & 92 & 5 & 4 & 1342221716 & 1342220865 \\
\object{IC4687} & 18h13m39.63s & $-$57d43m31.3s & 11.1 & 73 & 10 & 3 & 1342192993 & 1342204955 \\
\object{IRAS F18293-3413} & 18h32m41.13s & $-$34d11m27.5s & 11.5 & 78 & 11 & 2 & 1342192830 & 1342204954 \\
\object{IC4734} & 18h38m25.60s & $-$57d29m25.1s & 11.0 & 67 & 8 & 4 & 1342240013 & 1342229222 \\
\object{NGC6701} & 18h43m12.56s & +60d39m11.3s & 10.9 & 62 & 9 & 4 & 1342231994 & 1342229137 \\
\object{IRAS 19254-7245}\tablenotemark{a} & 19h31m20.50s & $-$72d39m21.8s & 11.8 & 270 & 7 & 2 & 1342231039 & 1342206210 \\
\object{IRAS 19297-0406} & 19h32m22.00s & $-$04d00m02.0s & 12.2 & 387 & 5 & 7 & 1342231078 & 1342230837 \\
\object{ESO339-G011} & 19h57m37.59s & $-$37d56m08.5s & 10.8 & 82 & 3 & 8 & 1342231990 & 1342230821 \\
\object{3C 405} & 19h59m28.36s & +40d44m01.9s &  & 252 & 2 & 7 & 1342246994 & 1342230853 \\
\object{IRAS 20087-0308} & 20h11m23.87s & $-$02d59m50.7s & 12.2 & 480 & 7 & 3 & 1342231049 & 1342230838 \\
\object{IRAS 20100-4156} & 20h13m29.54s & $-$41d47m34.9s & 12.4 & 595 & 6 & 4 & 1342245106 & 1342230817 \\
\object{MCG+04-48-002a} & 20h28m35.02s & +25d44m00.6s & 10.9 & 65 & 4 & 6 & 1342221682 & 1342233320 \\
\object{NGC6926} & 20h33m06.08s & $-$02d01m38.7s & 11.0 & 87 & 4 & 7 & 1342231050 & 1342218992 \\
\object{NGC6946} & 20h34m52.30s & +60d09m14.0s & 9.8 & 5 & 11 & 1 & 1342243603 & 1342188786 \\
\object{NGC6946\_05} & 20h35m12.01s & +60d08m55.2s & 9.8 & 5 & 4 & 8 & 1342224769 & 1342188786 \\
\object{IRAS 20414-1651} & 20h44m18.21s & $-$16d40m16.2s & 12.0 & 392 & 3 & 8 & 1342243623 & 1342231345 \\
\object{3C 424} & 20h48m12.03s & +07d01m17.5s &  & 586 & 0 & 9 & 1342255797 & 1342244149 \\
\object{IC 5063}\tablenotemark{a} & 20h52m02.10s & $-$57d04m06.6s & 10.2 & 46 & 3 & 8 & 1342242619 & 1342206208 \\
\object{CGCG448-020}\tablenotemark{a} & 20h57m24.33s & +17d07m38.3s & 11.7 & 161 & 10 & 2 & 1342221679 & 1342233327 \\
\object{ESO286-IG019} & 20h58m26.79s & $-$42d39m00.6s & 11.8 & 185 & 11 & 1 & 1342245107 & 1342230815 \\
\object{ESO286-G035} & 21h04m11.13s & $-$43d35m34.1s & 10.8 & 73 & 7 & 5 & 1342216901 & 1342230813 \\
\object{3C 433} & 21h23m44.60s & +25d04m27.1s &  & 465 & 0 & 9 & 1342245864 & 1342234675 \\
\object{NGC7130} & 21h48m19.50s & $-$34d57m04.7s & 11.1 & 69 & 12 & 1 & 1342219565 & 1342210527 \\
\object{NGC7172} & 22h02m01.91s & $-$31d52m11.3s & 10.2 & 36 & 4 & 7 & 1342219549 & 1342209301 \\
\object{ESO467-G027} & 22h14m39.85s & $-$27d27m50.5s & 10.8 & 74 & 2 & 8 & 1342245108 & 1342245428 \\
\object{IC5179} & 22h16m09.13s & $-$36d50m36.6s & 10.9 & 48 & 6 & 4 & 1342245109 & 1342244158 \\
\object{NGC7331} & 22h37m04.10s & +34d24m56.0s & 10.3 & 15 & 6 & 4 & 1342245871 & 1342189532 \\
\object{UGC12150} & 22h41m12.19s & +34d14m56.2s & 11.1 & 96 & 6 & 5 & 1342221699 & 1342220870 \\
\object{IRAS 22491-1808} & 22h51m49.26s & $-$17d52m23.5s & 11.9 & 345 & 9 & 2 & 1342245082 & 1342234671 \\
\object{NGC7465} & 23h02m00.96s & +15d57m53.4s & 9.7 & 30 & 3 & 8 & 1342245869 & 1342234763 \\
\object{NGC7469} & 23h03m15.62s & +08d52m26.4s & 11.3 & 72 & 13 & 0 & 1342199252 & 1342196915 \\
\object{ESO148-IG002} & 23h15m46.72s & $-$59d03m15.1s & 11.8 & 193 & 11 & 1 & 1342245110 & 1342209299 \\
\object{IC5298} & 23h16m00.64s & +25d33m23.7s & 11.3 & 122 & 7 & 5 & 1342221700 & 1342234766 \\
\object{NGC7552} & 23h16m10.77s & $-$42d35m05.4s & 10.7 & 21 & 12 & 1 & 1342198428 & 1342210528 \\
\object{NGC7591} & 23h18m16.26s & +06d35m08.8s & 10.8 & 72 & 6 & 7 & 1342257346 & 1342234758 \\
\object{NGC7592}\tablenotemark{a} & 23h18m22.08s & $-$04d24m57.6s & 11.1 & 106 & 7 & 4 & 1342221702 & 1342234750 \\
\object{NGC7582} & 23h18m23.50s & $-$42d22m14.0s & 10.5 & 21 & 11 & 2 & 1342209280 & 1342210529 \\
\object{IRAS 23230-6926} & 23h26m03.62s & $-$69d10m18.8s & 12.1 & 482 & 6 & 4 & 1342246276 & 1342230806 \\
\object{NGC7674} & 23h27m56.68s & +08d46m43.6s & 11.2 & 130 & 5 & 7 & 1342245858 & 1342234929 \\
\object{IRAS 23253-5415} & 23h28m06.10s & $-$53d58m31.0s & 12.1 & 595 & 4 & 6 & 1342246277 & 1342234737 \\
\object{NGC7679a}\tablenotemark{a} & 23h28m46.61s & +03d30m41.8s & 10.8 & 75 & 6 & 7 & 1342221701 & 1342234755 \\
\object{IRAS 23365+3604} & 23h39m01.27s\ & +36d21m08.7s & 12.0 & 290 & 7 & 5 & 1342224768 & 1342234919 \\
\object{NGC7771} & 23h51m24.88s & +20d06m42.6s & 11.1 & 63 & 10 & 3 & 1342212317 & 1342199379 \\
\object{Mrk331} & 23h51m26.80s & +20d35m09.9s & 11.2 & 81 & 12 & 0 & 1342212316 & 1342234682g
\enddata
\tablecomments{$n_{\\rm det}$ and $n_{\rm ul}$ indicate the number of 3 sigma detections and upper limits, respectively, reported in Table \ref{tbl:ftsflux} out of our 13 fitted lines: CO \jfour\ to \jthirteen, two \ci\ lines, and one \nii.}
\tablenotetext{a}{Indicates photometry correction was not performed on extended FTS spectrum, see Section \ref{sec:obs:sourcebeam}.}
\end{deluxetable*}

\subsection{Herschel SPIRE Photometry and FTS Spectra}\label{sec:obs:spectra}

All spectra used in the sample were reprocessed with HIPE\footnote{HCSS, HSpot, and HIPE are joint developments by the 
Herschel Science Ground Segment Consortium, consisting of ESA, the NASA Herschel Science Center, and the HIFI, PACS and SPIRE consortia.}  
 developer's version 13.0.3849 and \texttt{spire\_cal\_13\_0}, obtained from Rosalind Hopwood on 2014 September 30.  
This calibration corrects for rapidly changing telescope temperatures near the beginning of observation cycles, which has the largest effect on faint sources \citep{Swinyard2014}.  
Overall the calibration errors, even from earlier calibration sets, are within 6\% for point sources and 7\% for extended sources. 
The majority of the observations were done in sparse sampling mode, for which we took the spectra from the central pixels (SLWC3, SSWD4). 
For those mapping observations in intermediate or fully sampled mode, we extracted the spectrum from the pixel corresponding to the central coordinates of the map (those in Table \ref{tbl:galaxies}).
All SPIRE photometry was downloaded in 2014 September from the Herschel Science Archive (SPG v11.1.0) and not reprocessed.

\subsection{Source/Beam Correction}\label{sec:obs:sourcebeam}

The majority of the sources in our band are quite point-like compared to the SPIRE FTS beam, which varies from $\sim 45$\as\ to 17\as.  
Because the beam size is discontinuous between the upper frequency end of the SLW band and the lower frequency end of the SSW band, 
galaxies which are not point-like will show a notable discontinuity at this juncture. Even galaxies that are relatively small compared to the beam, but still 
not perfectly point-like, will show this discontinuity and require a correction to properly compare the emission across the SPIRE bandpass.
This is necessary because we only use the central FTS detectors.
An example is shown in Appendix \ref{sec:linefitexample}.

We perform the same source/beam correction as described in full detail in Section 2.2 of K14.  Briefly, we use the SPIRE Photometer Short Wave (PSW, 250 $\mu$m) maps, 
observed with 19\farcs32 beams, and convolve them to larger beam sizes ($\Omega_b$) and measure the new peak flux density. 
We compare this flux density to that of a $b$ = 43\farcs5 beam, which 
corresponds to the beam at the CO \jfour\ transition.
The ratio of the two flux densities, $\eta_{b, 43.5}$, as a function of beam size is between that expected for a point source (1) and a fully extended source ($\Omega_b/\Omega_{43.5}$).  
For each galaxy's unique distribution, for any beam size, we have a value of $\eta_{b, 43.5}$ to refer the emission to a 43\farcs5 beam.
We divide the SPIRE spectrum by $\eta_{b, 43.5}$ to refer the flux density at all wavelengths to that observed by a 43\farcs5 beam. 
We also use these values to refer CO integrated flux values measured from other facilities with smaller beam sizes to the 43\farcs5 beam (Section \ref{sec:obs:lowj}).

We apply an additional correction (also used in K14) to match the total flux density of the spectrometer with the photometer flux density. 
At the high frequency end, we match the total SSW flux density integrated over the photometer PSW bandpass, $\hat{F}(PSW)$, to that of the PSW photometer-integrated flux density at 43\farcs5, $F^{\prime \prime}(PSW)$ 
by multiplying the spectrum by $X_{\rm SSW}$ = $F^{\prime \prime}$(PSW)/$\hat{F}$(PSW).
There are two photometer bands (PMW and PLW) which overlap with the SLW band, so we define a line that connects those two ratios, $F^{\prime \prime}$(PMW)/$\hat{F}$(PMW) and $F^{\prime \prime}$(PLW)/ $\hat{F}$(PLW), and multiply the spectrum by that ratio as a function of wavelength, $X_{\rm SLW} (\nu)$.
This photometry correction step is often most significant in the SSW, which can overestimate the measured flux compared to the photometry.
Some spectra had somewhat over-subtracted telescope emission, giving slightly negative flux densities, especially at the lowest-frequency end.  In these cases, no correction was done to the SLW band to match the PLW photometry, which would use negative ratios.  In a few cases, we also did not correct the SLW band if the ratios derived from the PLW and PMW were significantly discrepant 
(i.e. would produce a non-sensical SLW continuum). 
The spectra that were not corrected are marked in Table \ref{tbl:galaxies}.

In order to compare the CO emission to \lfir, we must also correct \lfir\ to properly represent 
the same amount of emission as the CO within our beam. 
Similar to the procedure above, we convolve the SPIRE photometer maps at wavelength $\lambda$ to the beamsize of 43\farcs5 and find the ratio of the peak flux density in 
Jy measured with such a beam ($F_{\rm beam, \lambda}$) to the total integrated emission in the map ($F_{\rm total, \lambda}$). Assuming \lfir$_{\rm ,beam}$ = \lfir$_{\rm ,total}$ $\times F_{\rm beam, 250 \mu m} / F_{\rm total, 250 \mu m}$, and likewise for the 350 and 500 $\mu m$ maps, we can determine the proper \lfir$_{\rm ,beam}$ for comparison to the CO emission. 
The three photometers agreed well, and so we use the average of the $F_{\rm beam, \lambda} / F_{\rm total, \lambda}$ ratios.
Of the 232 observations with known redshift and available spectra, 118 have ratios of $< 0.8$, and 42 have ratios of $< 0.5$. We propagate the errors from the total measured integrated flux density through to the final measurement of \lfir$_{\rm ,beam}$.

\subsection{Herschel FTS Line Fitting Procedure}

The CO \jfour\ to \jthirteen\ lines, both \ci\ lines, and the \nii\ 205 $\mu$m line are the brightest lines in the FTS spectra. 
To fit these, we start with the FTFitter code from the University of Lethbridge.\footnote{\url{https://www.uleth.ca/phy/naylor/index.php?page=ftfitter}}  
Treating each detector (SLW and SSW) separately, the code fits a polynomial to the baseline, and then simultaneously fits 
unresolved lines at the expected frequencies of the lines listed above (given known redshifts).
We place a lower limit of the total area of the line profile to be above 0; we do not expect any of these lines to be in absorption.  
We limit the line center to within $\pm$ 500 \kms\ of that expected from the redshift to allow for uncertainty in the velocity scale and 
physical differences in the gas kinematics.  
In wavenumbers, this is about 0.025 - 0.084 cm$^{-1}$ over the band, compared to the FWHM of the line profile of 0.048 cm$^{-1}$.

We manually inspected the resulting fits to determine if any lines were clearly resolved. This is most likely to be case for the \nii\ line 
as velocity resolution is the highest at the higher frequencies, and it is much brighter than CO lines at similarly high frequencies which may be undetected.  
Resolved lines do not show the same characteristic ringing of the sinc function; 
the ringing is significantly lower, if not imperceptible, smeared out by the effective convolution of the emission line profile and instrumental profile.
We refit the lines that met this criteria as a Gaussian convolved with the instrumental line profile. In this case, the lines are barely resolved, thus Gaussian 
profiles are perfectly adequate (no more detailed velocity profiles can be determined from the FTS). 

The fact that SPIRE utilizes a Fourier Transform Spectrometer introduces a special problem in the treatment of line fitting.  The true measured 
quantity is the interferogram, or the interference pattern at the focus as a function of optical path difference (OPD) as the mirror of the interferometer moves linearly.
The spectrum itself is the Fourier Transform (FT) of this interferogram, which leads to two important consequences: 1) the wavelength bins are not truly independent, 
which many fitting routines assume, and 2) the resulting noise pattern closely resembles the FWHM = 0.048 cm$^{-1}$ sinc function line profile.  The result is that the errors 
output by a least-squares fitter, like the FTFitter and the built-in HIPE Spectrum Fitter routines, may not be an accurate representation of the line flux uncertainty.  Moreover, 
it can do an excellent job of fitting a ``ripple" in the spectrum which, to the observer's eye, may not be particularly distinguishable from any other ripple nearby, other than that we expect 
the e.g., CO line to correspond to the fitted ripple's wavelength, and no similarly strong lines to be adjacent in the spectrum.

Though the ideal situation would be to the fit the interferogram itself, this is not a user-accessible option for SPIRE data considering the many calibration steps 
that occur in processing after the FT.  Instead, we created a Bayesian analysis method to determine the probability distribution function of the true line flux 
given the observed line flux, $P(f_{true} | f_{obs})$, which is heavily influenced by the correlated noise pattern in the spectrum.  The noise itself is difficult to accurately 
characterize, varying from observation to observation, and across the bandpass of a given observation. 
Therefore, instead of attempting to describe the correlated noise for our entire sample, we focus on the area around each individual (unresolved) line.

We describe the procedure briefly here, but show a more in-depth example with illustrative figures for NGC4388 in Appendix \ref{sec:linefitexample}. This procedure is not used 
for lines that were manually identified as resolved, which are already high signal/noise. For each line, 
we input sinc profile lines of varying amplitudes $f_{true}$ over the region $\pm$ 2 cm $^{-1}$ from the line center (excluding the area immediately around any CO, \ci, or \nii\ lines) 
and then refit the spectrum. We compare the measured integrated fluxes, $f_{obs}$ to the known input values, $f_{true}$. The probability distribution function for our CO line is the distribution of 
input fluxes that produced that particular measured flux value, a slice of the $P(f_{obs} | f_{true})$ two-dimensional distribution that we created.

For high signal-to-noise line detections, the procedure replicates a Gaussian distribution of similar median and error ($\sigma$) as the parameters estimated by the FTFitter. This is because a very high amplitude line input, added 
anywhere in the spectrum, will return the same integrated flux value as we input ($f_{obs}$ = $f_{true}$). However, a line with smaller amplitude may add constructively or destructively to the underlying ripple pattern of the spectrum, returning a higher or lower flux than input. A local variation from the detector's average baseline may also influence the final fitted value, which may shift the median value of the probability distribution function of the line flux.

The procedure makes the most difference for the high-$J$ CO lines; 60\% of the 3$\sigma$ detections of CO \jthirteen\ from least-squares fitting were shown to 
have $>3\sigma$ uncertainty. For all the CO lines in the SSW band ($J_{\rm upper} \ge 9$), this number is 44\%. For the CO lines in the SLW band, it is only 15\%. The numbers are the lowest for \nii\ (4\%) because it is so bright, and for \ci\ \jtwo\ and CO \jseven\ (5\%, 8\%) because they lie in the lowest noise part of the spectrum and are relatively bright.

The results from this line fitting procedure are shown in Table \ref{tbl:ftsflux}. The median, $-1\sigma$, and $+1\sigma$ values are derived from the points at which the cumulative distribution functions (CDF) equal 0.5, 0.159, and 0.841, respectively. If $-1\sigma$/median is less than 3, a value for a $3\sigma$ upper limit is also shown (where the CDF = 0.997).

\begin{deluxetable*}{lrccccc}
\tablecaption{Line Fluxes and Uncertainty Ranges from SPIRE FTS\label{tbl:ftsflux}}
\tablehead{\colhead{Galaxy} & \colhead{Line} & \colhead{Resolved?} & \colhead{Median} & \colhead{-1$\sigma$} & \colhead{+1$\sigma$} & \colhead{3$\sigma$ Upper Limit}\\ \colhead{} & \colhead{} & \colhead{} & \colhead{Jy km s$^{-1}$} & \colhead{Jy km s$^{-1}$} & \colhead{Jy km s$^{-1}$} & \colhead{Jy km s$^{-1}$}}
\startdata
NGC0023 & CI1-0 &  & 2.57e+02 & 6.81e+01 & 5.57e+02 & 7.98e+02 \\
NGC0023 & CI2-1 &  & 3.82e+02 & 3.09e+02 & 4.57e+02 &  \\
NGC0023 & CO4-3 &  & 3.12e+02 & 7.15e+01 & 6.10e+02 & 1.21e+03 \\
NGC0023 & CO5-4 &  & 7.38e+02 & 5.76e+02 & 9.69e+02 &  \\
NGC0023 & CO6-5 &  & 4.23e+02 & 3.26e+02 & 4.88e+02 &  \\
NGC0023 & CO7-6 &  & 3.24e+02 & 2.42e+02 & 3.97e+02 &  \\
NGC0023 & CO8-7 &  & 2.49e+02 & 1.18e+02 & 3.83e+02 & 5.71e+02 \\
NGC0023 & CO9-8 &  & 1.15e+02 & 2.51e+01 & 2.94e+02 & 5.63e+02 \\
NGC0023 & CO10-9 &  & 2.22e+02 & 5.53e+01 & 3.56e+02 & 5.28e+02 \\
NGC0023 & CO12-11 &  & 7.78e+01 & 1.81e+01 & 1.92e+02 & 3.26e+02 \\
NGC0023 & CO13-12 &  & 2.29e+02 & 9.69e+01 & 3.55e+02 & 4.76e+02 \\
NGC0023 & NII & X & 2.63e+03 & 2.46e+03 & 2.80e+03 &  \\
NGC34 & CI1-0 &  & 4.96e+02 & 3.36e+02 & 6.45e+02 &  \\
NGC34 & CI2-1 &  & 4.29e+02 & 3.83e+02 & 4.72e+02 &  \\
NGC34 & CO5-4 &  & 7.87e+02 & 6.92e+02 & 8.77e+02 & 
\enddata
\tablecomments{Table \ref{tbl:ftsflux} is published in its entirety in the electronic edition of The Astrophysical Journal. A portion is shown here for guidance regarding its form and content.}
\end{deluxetable*}

\subsection{Low-$J$ CO Lines from the Literature and Arizona Radio Observatory}\label{sec:obs:lowj}

The bandpass of the {\it Herschel} FTS starts around the CO \jfour\ line, but the majority of the molecular mass in galaxies is cool and populates the lower rotational levels.  
We complement the line fluxes derived from the FTS with the CO \jone, \jtwo, and \jthree\ lines available from ground-based observatories.  
Many of these galaxies have already been studied in the literature, especially in large CO surveys.  

\begin{deluxetable*}{lrrrrcrrrrr}
\tablecaption{CO \jone\ to \jthree\ Line Fluxes\label{tbl:lowjflux}}
\tablehead{\colhead{Galaxy} & \colhead{$J_{\rm up}$} & \colhead{Reported} & \colhead{$\sigma_m$} & \colhead{$\sigma_c$} & \colhead{Unit} & \colhead{$\Delta_{v}$} & \colhead{$\Omega_{b}$} & \colhead{$I \Delta_{v}$ ($\Omega_{FTS}$)} & \colhead{$\sigma$} & \colhead{Ref}\\ \colhead{ } & \colhead{ } & \colhead{ } & \colhead{ } & \colhead{ } & \colhead{$\times$\kms} & \colhead{ } & \colhead{ } & \colhead{ } & \colhead{ } & \colhead{ }}
\startdata
NGC0023 & 1 & 16.9 &  & 4.2 & K & 141 & 24 & 138.9 & 34.7 & 1 \\
NGC0023 & 1 & 6.0 &  & 1.2 & K & 374 & 55 & 169.9 & 34.0 & 2 \\
NGC0023 & 1 & 8.9 &  & 2.2 & K & 190 & 45 & 184.4 & 46.1 & 1 \\
NGC0023 & 1 & 18.0 &  & 1.3 & K &  & 33 & 234.0 & 16.9 & 3 \\
NGC0023 & 1 & 34.0 & 0.4 & 6.8 & K &  & 22 & 247.4 & 49.6 & 4 \\
NGC0023 & 2 & 18.8 &  & 4.7 & K & 129 & 12 & 235.6 & 58.9 & 1 \\
NGC0023 & 2 & 7.4 &  & 1.9 & K & 210 & 24 & 243.2 & 60.8 & 1 \\
NGC0023 & 3 & 15.3 & 1.3 & 2.3 & K & 257 & 22 & 1003.0 & 171.8 & 5 \\
NGC34 & 1 & 17.0 &  & 4.2 & K & 149 & 24 & 115.3 & 28.8 & 1 \\
NGC34 & 1 & 4.5 & 0.3 & 0.4 & K & 295 & 55 & 131.8 & 16.0 & 6 \\
NGC34 & 1 & 6.7 &  & 1.7 & K & 274 & 45 & 138.4 & 34.6 & 1 \\
NGC34 & 1 & 148.5 & 13.5 & 29.7 & Jy &  & 45 & 147.6 & 32.4 & 7 \\
NGC34 & 2 & 4.3 &  & 1.1 & K & 271 & 24 & 116.7 & 29.2 & 1 \\
NGC34 & 2 & 56.5 &  & 14.1 & K & 172 & 12 & 471.9 & 118.0 & 1 \\
NGC34 & 3 & 7.7 & 1.8 & 1.2 & K & 168 & 22 & 404.8 & 110.5 & 5
\enddata
\tablecomments{Table \ref{tbl:lowjflux} is published in its entirety in the electronic edition of The Astrophysical Journal. A portion is shown here for guidance regarding its form and content. 
{\bf The first eight columns refer specifically to the measurements reported in the literature. ``Reported" is the reported value in the units of the ``Units" column.}
$\sigma_m$ and $\sigma_c$ refer to measurement and calibration error, if separately reported, otherwise calibration errors are assumed or contain total error. $\Delta_v$ is the FWHM of the line, if reported, and $\Omega_b$ is the FWHM of the beam size. {\bf The next two columns are the values used in our analysis:} $I\Delta_v (\Omega_{FTS})$ is the flux in Jy km/s, referred to the 43\farcs5 beam, and $\sigma$ is the total accompanying error. 
{\bf References.} (1) \citet{Albrecht2007}; (2) \citet{Sanders1991}; (3) \citet{Elfhag1996}; (4) \citet{Garcia-Burillo2012}; (5) SMT (this work); (6) \citet{Maiolino1997}; (7) \citet{Baan2008}; (8) 12M (this work); (9) \citet{Leroy2006}; (10) \citet{Bayet2006}; (11) \citet{Mao2010}; (12) \citet{Mirabel1990}; (13) \citet{Garay1993}; (14) \citet{Harrison1999}; (15) \citet{Young1995}; {\bf (16) \citet{Earle2008}}; (17) \citet{Solomon1997}; (18) \citet{Papadopoulos2012}; (19) \citet{Evans2005}; (20) \citet{Aalto1995}; (21) \citet{Leech2010}; (22) \citet{Kamenetzky2011}; (23) \citet{Spinoglio2012}; (24) \citet{Young2011}; (25) \citet{Alatalo2011}; (26) \citet{Lazareff1989}; (27) \citet{Papadopoulos1998}; (28) \citet{Sandqvist1995}; (29) \citet{Sandqvist1999}; (30) \citet{Bothwell2013}; (31) \citet{Ward2003}; (32) \citet{Yao2003}; (33) \citet{Sliwa2012}; (34) \citet{Boselli2014}; (35) \citet{Schirm2014}; (36) \citet{Wild2000}; (37) \citet{Eckart1990}; (38) \citet{Mauersberger1999}; (39) \citet{Greve2009}; (40) \citet{Claussen1992}.}
\end{deluxetable*}

For some galaxies, we also performed single-dish measurements using the Arizona Radio Observatory (ARO). 
Measurements of the CO \jone\ line were conducted with the 12m dish on Kitt Peak in May of 2015, and those of CO \jtwo\ and \jthree\ were conducted with 
the Submillimeter Telescope (SMT) located on Mt. Graham from November 2014 to February 2015.
At the 12m, we used the ALMA Type 3 mm receiver with two 2 MHz backends in series, yielding 2.6 \kms\ channel resolution and about 670 \kms\ bandwidth.
At the SMT, we used the ALMA Type 1.3 mm sideband separating receiver (for CO \jtwo) and the 0.8 mm double sideband receiver (for CO \jthree) 
with the 1 MHz filterbanks in 2IF mode. Most observations were conducted with beam switching, except for highly extended sources which required position switching. 
Pointing and focus were checked periodically on planets or bright continuum sources. 

Beam efficiency measurements were conducted with Venus and Jupiter\footnote{Additional planetary observations provided by Karin Sandstrom.} using the procedure described in \citet{Schlingman2011}.
For CO \jone, \jtwo, and \jthree\ we found $\eta_{\rm mb}$ of 0.86-0.89, 0.57-0.62, and 0.58-0.65, respectively. 
The beam sizes are approximately 55\as, 32\as, and 22\as, respectively.
The spectra were reduced, baseline subtracted, and converted to $T_{\rm mb}$ scale using 
$\eta_{\rm mb}$ in CLASS. 
Finalized spectra were smoothed to approximately 20 \kms\ bins 
from which total integrated fluxes were derived. 

All low-J lines utilized in this work, including the ones from ARO, are available in Table \ref{tbl:lowjflux}.
As Table \ref{tbl:lowjflux} shows, the low-$J$ lines we use come from a variety of telescopes with different beam sizes.
For subsequent comparison to Herschel CO lines, all line fluxes are referenced to the 43\farcs5 beam size using the same ratios ($\eta_{b,43.5}$) as described in Section \ref{sec:obs:sourcebeam}. 
Table \ref{tbl:lowjflux} lists both the literature reported values (third column, in their original beam sizes and units) and the 43\farcs5 referenced fluxes in Jy \kms\ we use in our analysis (ninth column).

\section{Analysis}\label{sec:analysis}

We examined the relationship between \lprimeco\footnote{\lprimeco = 3.25 $\times 10^7 S_{\rm CO} \Delta v D_L^2 (1+z)^{-3} \nu_{obs}^{-2}$ [K \kms pc$^2$], where $D_L$ is the luminosity distance in Mpc, 
$\nu_{obs}$ in GHz, $S_{\rm CO} \Delta v$ in Jy \kms, from \citet{Carilli2013, Solomon1992}. Note that \lprimeco\ is just the area within the beam times the velocity-integrated antenna temperature, $A \times T \Delta v$, where A is the area $10^{12} \Omega D_L^2 (1+z)^{-4}$ [pc$^2$].}\
 and \lfir\ (from 40-120 $\mu$m),  similar to \citet{Lu2014}, \citet{Greve2014} and \citet{Liu2015}, discussed further in Section \ref{sec:analysis:comparison}. 
 We choose the orientation of our axes, with \lfir\ on the y-axis, for the easiest comparison to the existing literature. This orientation comes from the comparison to the Kennicutt-Schmidt (K-S) scaling law, which relates star formation rate (SFR) to molecular gas surface density, with $\dot{\Sigma}_* \propto \Sigma_{\rm gas}^{1.4 \pm 0.15}$ \citep{Kennicutt1998}. Subsequent large-scale analyses have found indices from 1.0-1.4 for the molecular gas and higher values, 1.4-3.1, for the total gas surface density \citep{Kennicutt2012}.
 \lfir\ can be considered a proxy for SFR if one excludes the contribution of AGN to \lfir, which we admittedly do not separate here. It is likely small for most sub-ULIRG galaxies, especially at these wavelengths. In the case of \jone, the well-known ``X-factor" or $\alpha_{\rm CO}$ is used to relate the luminosity of CO \jone\ to the total molecular mass, so the slope derived here is comparable to the K-S relation. 
 However, this relationship is not applicable at higher-$J$, where the CO luminosity is not a tracer of mass (K14), but we choose to keep the same orientation to avoid confusion. Neither variable should be considered the ``independent" one, regardless of which appears on the x-axis.

The theoretical explanation for these relationships were first investigated to describe the discrepant power laws in low-$J$ emission of various molecules, namely CO \jone, where \lfir $\propto$ \lco$^{1.4-1.6}$, and HCN \jone, where \lfir $\propto L_{\rm HCN}^{1.0}$ \citep{Gao2004a, Gao2004b}. The CO slope closely resembles that of the aforementioned K-S relation slope, but HCN does not match.
\citet{Krumholz2007} showed that this could be understood as a consequence of the different critical densities for different species' ground-state transitions, assuming isothermality.
In short, a molecular line with a low critical density compared to a galaxy's median gas density, $\rho_g$, will be excited by the majority of the molecular gas. The star formation rate will therefore depend on one factor of density $\rho_g$ based on the total amount available for star formation, and a factor $\rho_g^{0.5}$ from the dynamical timescale of the gas, adding up to a total factor of 1.5 in the case of the low-$n_{crit}$ CO \jone\ line.  On the other hand, when the molecular line has a critical density higher than that of the median gas density, its emission will be picked out from high-density peaks only,  specifically peaks of the same density (and therefore same free-fall time) across different galaxies. Therefore the higher $n_{crit}$ of HCN \jone\ yields a slope of 1.0. We discuss our results in context of these expectations in Section \ref{sec:disc}.

\begin{figure*} 
\centering
\includegraphics[width=7in]{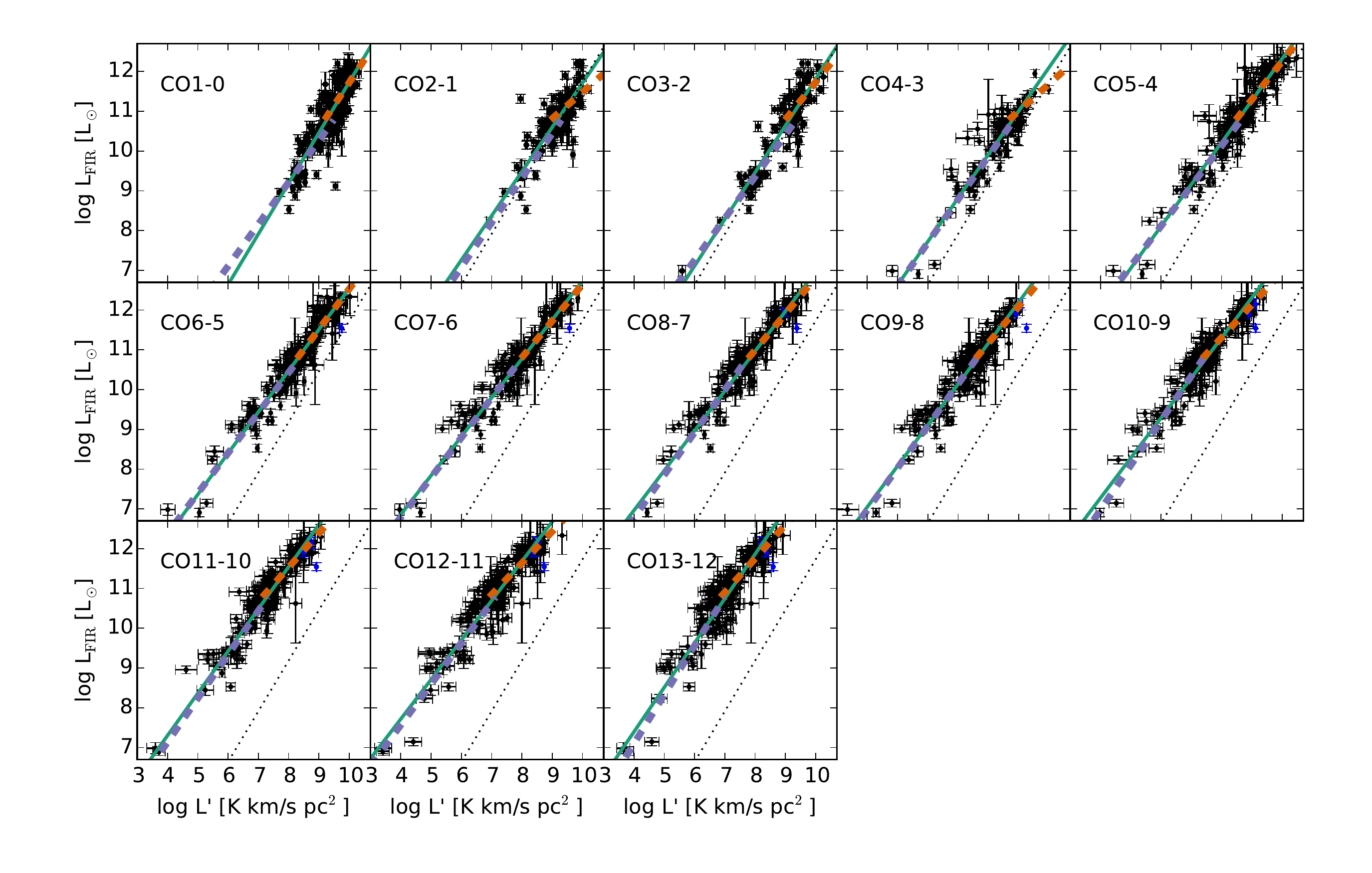} 
\caption[CO vs. \lfir]{CO vs. \lfir. The y-axis is the \lfir\ in the beam for comparison to the CO measurement (see Section \ref{sec:obs:sourcebeam}. Low-$J$ lines may include multiple measurements for the same galaxy if available in the literature (Table \ref{tbl:lowjflux}). Blue data points indicate resolved (Sinc-Gaussian) measurements from the FTS. Line fits are described in Section \ref{sec:analysis}. The green line is fitting the whole sample; the CO \jone\ line fit is shown as a dotted black line on each other CO plot for comparison. The dashed orange and purple lines are fits when separating the sample into galaxies above and below \lfir = $6 \times 10^{10}$ \ls, respectively.
\label{fig:slopes}}
\end{figure*}

\subsection{Fitting \lprimeco/\lfir\ Slopes}\label{sec:analysis:slopes}

As mentioned in Section \ref{sec:obs:sourcebeam}, all fluxes including \lfir\ are referred to the emission within a 43\farcs5 beam.
To determine the coefficients of the relation log(\lfir) = a log(\lprimeco) + b, we used the python module lnr.bces\footnote{\url{http://home.strw.leidenuniv.nl/~sifon/pycorner/bces/}, by Crist\'obal Sif\'on}, 
which utilizes the Bivariate, Correlated Errors and intrinstic Scatter method of \citet{Akritas1996}. This is important because we have errors on both variables (we introduced a non-negligible error into the \lfir\ variable through our source/beam correction). As stated above, we chose the examination of \lfir\ as a function of \lprimeco\ to match the most recent literature and note that the solution to the inverse problem 
(\lprimeco\ as a function of \lfir) does not simply produce best-fit slopes that are the inverse of those presented here. For the case of low-$J$ lines collected from the literature and our ARO follow-up, multiple measurements for the same galaxies (positioned at the location matching the FTS coordinates) are treated independently.

\begin{deluxetable}{rrrrrr}
\tablecaption{Correlations between \lprimeco\ and \lfir: Full Sample.\label{tbl:slopes}}
\tablehead{\colhead{$J_{\rm up}$} & \colhead{a} & \colhead{$\sigma_a$} & \colhead{b} & \colhead{$\sigma_b$} & \colhead{n}}
\startdata
1 & 1.27 & 0.04 & -1.0 & 0.4 & 299 \\
2 & 1.11 & 0.07 & 0.6 & 0.7 & 138 \\
3 & 1.18 & 0.03 & 0.1 & 0.3 & 131 \\
4 & 1.09 & 0.05 & 1.2 & 0.4 & 108 \\
5 & 1.05 & 0.03 & 1.8 & 0.3 & 195 \\
6 & 1.04 & 0.03 & 2.2 & 0.2 & 199 \\
7 & 0.98 & 0.03 & 2.9 & 0.2 & 196 \\
8 & 1.00 & 0.03 & 3.0 & 0.3 & 186 \\
9 & 1.03 & 0.04 & 2.9 & 0.3 & 176 \\
10 & 1.01 & 0.03 & 3.2 & 0.3 & 184 \\
11 & 1.06 & 0.04 & 3.1 & 0.3 & 166 \\
12 & 0.99 & 0.03 & 3.7 & 0.2 & 168 \\
13 & 1.12 & 0.04 & 2.9 & 0.3 & 156
\enddata
\tablecomments{Best fit measurements and errors for Log(\lfir)=a Log(\lprimeco)+b. Column n = number of data points used in relation (not all 3$\sigma$ detections).}
\end{deluxetable}

When including all spectra in our sample, we find slopes starting at $1.3$ for CO \jone, and lowering to about $\sim 1$ for the mid- to high-$J$ CO lines, with no discernible trend with increasing J. The results are shown in Table \ref{tbl:slopes}. 
There is no significant difference whether we include or exclude FTS lines with S/N $<$ 3.
We also separated our samples into a few categories that are somewhat overlapping, and summarize the differences here:

\begin{itemize}
\item We separated our galaxies into known AGN (categorized in Hyperleda\footnote{\url{http://leda.univ-lyon1.fr/leda/rawcat/a109.html/}} as quasar or any type of Seyfert) or not. 
This classification (using the \texttt{agnclass} category) is taken from the \citet{Veron-Cetty2006} catalog.
Within those classified by Hyperleda as AGN (78 of the 232 galaxies), 
no information is provided regarding the relative SF vs. AGN contributions to the total \lfir, so this is a somewhat crude division of the galaxy sample.
Looking only at AGN (Table \ref{tbl:slopesagn}), compared to the whole sample we find higher slopes for the low-$J$ lines ($1.5 \pm 0.1$, $1.2 \pm 0.1$, $1.3 \pm 0.1$, $1.1 \pm 0.1$, $1.15 \pm 0.04$ for \jone\ through \jfive), but the slope error bars overlap by CO \jsix\ and continue to follow the same trend as the whole sample. Looking at the sample that completely excludes AGN, we find a lower slope than the whole sample for CO \jone\ (1.13 $\pm$ 0.06), but at subsequently higher-J the slopes are not distinguishable from the combined sample. In summary, comparing AGN to the non-AGN sample, the most significant difference is in the CO \jone\ line. There are also differences, at less significance, up to the \jfive\ line. 
The low-J CO emission is not expected to be affected by the AGN; it is the \lfir\ and high-J CO likely being influenced. 
We likely do not see any difference because our AGN-designated galaxies (which are only about one third of the full sample) may not be entirely dominated in their molecular excitation from the AGN; as mentioned, we do not separate the relative SF vs. AGN contributions to total \lfir. Better quantification of the AGN influence, and higher spatial resolution, may result in differences in the slope.

\item Astronomers often separate galaxies into (U)LIRGs or lower luminosity galaxies. We separate our galaxies at \lfir\ = $6 \times 10^{10}$ \ls\, which we found is approximately equal to \lir\ (8-1000 $\mu$m) = $1 \times 10^{11}$, based on the luminosities listed in \citet{Greve2014} and \citet{Kamenetzky2014}. (The exact cutoff value does not change the conclusions that follow.) The CO \jone\ line has been known to not be fit by one slope among (U)LIRGs and lower luminosity galaxies; the superlinear slope that results from a single power line fit is due to higher dense gas fractions in (U)LIRGs \citep{Gao2004b, Greve2014}, essentially creating a higher intercept (but same slope). 
We do find lower slopes in \jone\ among these two populations fit separately than their combined slope of 1.3 $\pm$ 0.4.
We find the mid-J lines of CO are well fit by a single slope across many orders of magnitude; the slopes in all three cases (full sample, just (U)LIRGs, just galaxies with lower luminosities than (U)LIRGs) are not statistically distinguishable given the error bars. Our three highest-J CO lines (\jeleven, \jtwelve\, \jthirteen), however, do show a measurable difference in the best fit slope. Focusing primarily on (U)LIRGs changes the slopes at higher-$J$, decreasing to about 0.83 $\pm$ 0.03 (weighted average of three highest-J CO lines in Table \ref{tbl:slopesulirgs}), a highly significant difference from 1. The results for these two populations are shown in Tables \ref{tbl:slopesulirgs} and \ref{tbl:slopesnotlirgs}.

\item We restricted the fit to only those galaxies not particularly well resolved, those with \lfir$_{\rm ,beam} \ge 0.8$ \lfir$_{\rm ,total}$ based on the photometry maps, to see if our infrared luminosity correction could be influencing the y-axis values. We find lower slopes in this case, $0.88 \pm 0.06$ on average for high-J lines, but this may be due to the fact that 2/3 of this population are (U)LIRGs, not because large \lfir\ corrections are necessarily inaccurate.
\end{itemize}

Certainly, these slopes are masking a considerable amount of scatter in the data, and different trends can be discerned when fitting different populations (also seen in \citet{Liu2015}). The remaining figures of this paper examine the survey populations themselves with comparisons to these derived, ``global" relations.

\begin{deluxetable}{rrrrrr}
\tablecaption{Correlations between \lprimeco\ and \lfir: AGN Only.\label{tbl:slopesagn}}
\tablehead{\colhead{$J_{\rm up}$} & \colhead{a} & \colhead{$\sigma_a$} & \colhead{b} & \colhead{$\sigma_b$} & \colhead{n}}
\startdata
1 & 1.54 & 0.07 & -3.4 & 0.7 & 99 \\
2 & 1.24 & 0.14 & -0.6 & 1.2 & 39 \\
3 & 1.30 & 0.08 & -1.1 & 0.7 & 39 \\
4 & 1.15 & 0.08 & 0.6 & 0.6 & 33 \\
5 & 1.15 & 0.04 & 0.9 & 0.3 & 60 \\
6 & 1.09 & 0.04 & 1.6 & 0.4 & 62 \\
7 & 1.03 & 0.04 & 2.5 & 0.3 & 61 \\
8 & 1.03 & 0.05 & 2.6 & 0.4 & 58 \\
9 & 1.02 & 0.06 & 2.9 & 0.5 & 53 \\
10 & 1.01 & 0.05 & 3.1 & 0.4 & 58 \\
11 & 1.01 & 0.07 & 3.3 & 0.5 & 56 \\
12 & 0.99 & 0.06 & 3.6 & 0.4 & 55 \\
13 & 1.13 & 0.07 & 2.7 & 0.5 & 53
\enddata
\tablecomments{Best fit measurements and errors for Log(\lfir)=a Log(\lprimeco)+b. Column n = number of data points used in relation (not all 3$\sigma$ detections).}
\end{deluxetable}

\begin{deluxetable}{rrrrrr}
\tablecaption{Correlations between \lprimeco\ and \lfir: (U)LIRGs Only.\label{tbl:slopesulirgs}}
\tablehead{\colhead{$J_{\rm up}$} & \colhead{a} & \colhead{$\sigma_a$} & \colhead{b} & \colhead{$\sigma_b$} & \colhead{n}}
\startdata
1 & 1.15 & 0.09 & 0.2 & 0.8 & 225 \\
2 & 0.66 & 0.15 & 4.9 & 1.4 & 100 \\
3 & 0.94 & 0.11 & 2.3 & 1.1 & 86 \\
4 & 0.68 & 0.13 & 4.9 & 1.1 & 44 \\
5 & 0.96 & 0.07 & 2.7 & 0.6 & 129 \\
6 & 1.02 & 0.05 & 2.4 & 0.4 & 132 \\
7 & 0.92 & 0.04 & 3.5 & 0.4 & 131 \\
8 & 0.91 & 0.04 & 3.7 & 0.3 & 125 \\
9 & 0.92 & 0.07 & 3.8 & 0.5 & 111 \\
10 & 0.83 & 0.03 & 4.7 & 0.3 & 126 \\
11 & 0.86 & 0.05 & 4.7 & 0.4 & 115 \\
12 & 0.79 & 0.04 & 5.3 & 0.3 & 114 \\
13 & 0.85 & 0.05 & 5.0 & 0.4 & 109
\enddata
\tablecomments{Best fit measurements and errors for Log(\lfir)=a Log(\lprimeco)+b. Column n = number of data points used in relation (not all 3$\sigma$ detections).}
\end{deluxetable}

\begin{deluxetable}{rrrrrr}
\tablecaption{Correlations between \lprimeco\ and \lfir: Non-(U)LIRGs Only.\label{tbl:slopesnotlirgs}}
\tablehead{\colhead{$J_{\rm up}$} & \colhead{a} & \colhead{$\sigma_a$} & \colhead{b} & \colhead{$\sigma_b$} & \colhead{n}}
\startdata
1 & 1.05 & 0.09 & 0.8 & 0.8 & 74 \\
2 & 1.12 & 0.11 & 0.4 & 0.9 & 38 \\
3 & 1.05 & 0.05 & 0.9 & 0.4 & 45 \\
4 & 1.09 & 0.07 & 1.2 & 0.5 & 64 \\
5 & 1.01 & 0.06 & 2.1 & 0.5 & 66 \\
6 & 1.01 & 0.05 & 2.3 & 0.4 & 67 \\
7 & 1.00 & 0.06 & 2.8 & 0.4 & 65 \\
8 & 1.07 & 0.07 & 2.5 & 0.5 & 61 \\
9 & 1.06 & 0.08 & 2.7 & 0.5 & 65 \\
10 & 1.12 & 0.07 & 2.5 & 0.5 & 58 \\
11 & 1.10 & 0.09 & 2.7 & 0.6 & 51 \\
12 & 1.03 & 0.07 & 3.4 & 0.4 & 54 \\
13 & 1.23 & 0.08 & 2.2 & 0.5 & 47
\enddata
\tablecomments{Best fit measurements and errors for Log(\lfir)=a Log(\lprimeco)+b. Column n = number of data points used in relation (not all 3$\sigma$ detections).}
\end{deluxetable}

Fitting a linear relation in log space requires creating error bars that are symmetric in log space, which ours are not. The exaggeration of the linear errors when viewed in log space is highest for those line measurements with the lowest signal-to-noise ratio. 
To test whether the conversion to these symmetric error bars (which decreases the size of the lower error bar) introduces systematic bias in deriving the slopes, we chose fixed values of the slopes, set the nominal y-values to match the chosen slope, and draw  randomly from each data point's x and y error bars in linear space.
Fitting such scattered distributions many times results in distributions centered upon the chosen slopes and with scatter comparable to our error bars in Table \ref{tbl:slopes}. 
We find no evidence of systematic bias due to this effect.

We also investigated whether our error bars are responsible for the slightly lower value of the CO \jone\ line's slope at 1.3 instead of 1.4-1.6. This value is not due to the error bars in either the x- or y-directions; we still find a slope of about 1.3 if we exclude one, the other, or both in the line fitting. We do find a slope of 1.44 if we restrict the fit to those data points with log (\lprimeco) $>$ 9, indicating that the lower luminosity points are bringing down the slope overall. 

\subsection{Comparison to Similar Works}\label{sec:analysis:comparison}

\citet{Greve2014} conducted a comparison of CO emission to \lfir\ for the non-extended galaxies of the HerCULES sample \citep{vanderWerf2010} and \
found sharply decreasing slopes between log(\lfir) and log(\lprimeco), 
starting around 1 at CO \jone\ and decreasing as one moves to higher-$J$ starting around CO \jsix\ 
(0.93 $\pm$ 0.05, down to 0.47 $\pm 0.20$ by the \jthirteen\ line).

They interpret this sublinear slope as an indication that FUV radiation fields are not responsible for the the dense, warm gas emitting 
in the high-$J$ lines.
Mechanical heating and shocks are the likely explanation 
for the high-$J$ excitation, as was shown in many galaxies of the sample of K14 (and references therein). 
We are in agreement with \citet{Greve2014} that mechanical feedback can be related to star formation, such as stellar outflows, winds, and supernova remnant expansion; not being related to the FUV radiation from stars does not mean the excitation is not related to star formation at all.

The fluxes used in \citet{Greve2014} are reported in \citet{Rosenberg2015}. 
Their sample is included in ours, but our measured fluxes do not match in all cases. 
For lines up to \jten, 2/3 of their 18 non-extended galaxies match our flux values within 30\%. At higher-$J$, this number is more like 1/3. When discrepant, their values are often $\sim$ 50\% higher, or even factors of a few for IRASF05189-2524, for which we do not use the same spectra.
The major difference for other galaxies is likely the treatment of the source/beam correction. 
Importantly, \citet{Greve2014} only included (U)LIRGs (log(\lfir) $\ge$ 11).  As discussed above, we find lower slopes if we restrict ourselves to this population, but not as low as the 
slopes reported by \citet{Greve2014}, due to the differences in our measured fluxes.

\citet{Liu2015} conducted a larger study, more comparable to ours, utilizing the full extent of the {\it Herschel} archive. 
With the inclusion of galaxies from log(\lfir) $\ge$ 8, they found linear slopes between 
 log(\lfir) and log(\lprimeco) throughout the CO ladder starting at \jfour\ (statistically consistent with our results). Also consistent with our results and \citet{Greve2014}, 
 restricting the fits to the HerCULES sample of (U)LIRGs yields sublinear slopes for the highest-J lines.
 Their work does not include a table of the galaxies included in their sample or the 
 line fluxes used, so we cannot make a direct comparison of the fluxes used for our relations at this time. Some major differences may be important in the comparison of our work: 
 first, they use SPIRE calibration version 12 and we use version 13, which may make the most difference for galaxies observed near the start of SPIRE cooling cycles (due to ``cooler burp"). Second,
each of their CO and \lfir\ relations use a different beam size, that of the frequency of the CO line in the FTS. 
Third, for sparsely and intermediately sampled galaxies, 
they extract spectra from multiple bolometers in the detector array, meaning they are including multiple spatially Nyquist-sampled data points in their relationships (with their own matching \lfir\ values) for such galaxies.
Finally, their line fitting procedure uses sinc-convolved-gaussian (SCG) lines 
in HIPE for sparse observations and sinc for intermediate or full sampling observations. They note that SCG-derived fluxes are systematically higher, but 
we found that few galaxies have CO linewidths large enough to be resolved by the FTS.
Differences in the slopes of (U)LIRGs among \citet{Liu2015}, \citet{Greve2014}, and this work can also be attributed to the small dynamical range spanned by (U)LIRGs.
Despite these differences, our results agree well in finding mid- to high-$J$ CO slopes of 1 in a large sample of galaxies and less than 1 for (U)LIRGs.

\citet{Lu2014} also examined the CO \jfour\ through \jthirteen\ emission of the 65 LIRGs in the Great Observatories All-Sky LIRG Survey (GOALS), comparing to the IRAS 60-to-100 $\mu$m color, C(60/100), as a proxy for dust temperature. They demonstrated that \lir\ is not the best predictor of SLED shapes; we find an overall trend in Figure \ref{fig:averagesleds} (the line luminosity ratios relative to \jone\ in average SLEDs by \lfir\ bin increase with \lfir), but also a significant amount of variation in Figure \ref{fig:lineratios} (individual mid- to high-J CO/\jone\ luminosity ratios for each galaxy, unbinned), discussed in the next section. C(60/100), which is not presented here, is potentially a better predictor of the CO SLED shape (based on the line at which the SLED peaks in luminosity).

\subsection{Average SLEDs by \lfir\ Ranges}\label{sec:disc:averagesleds}

\begin{figure}  
\centering
\includegraphics[width=\columnwidth]{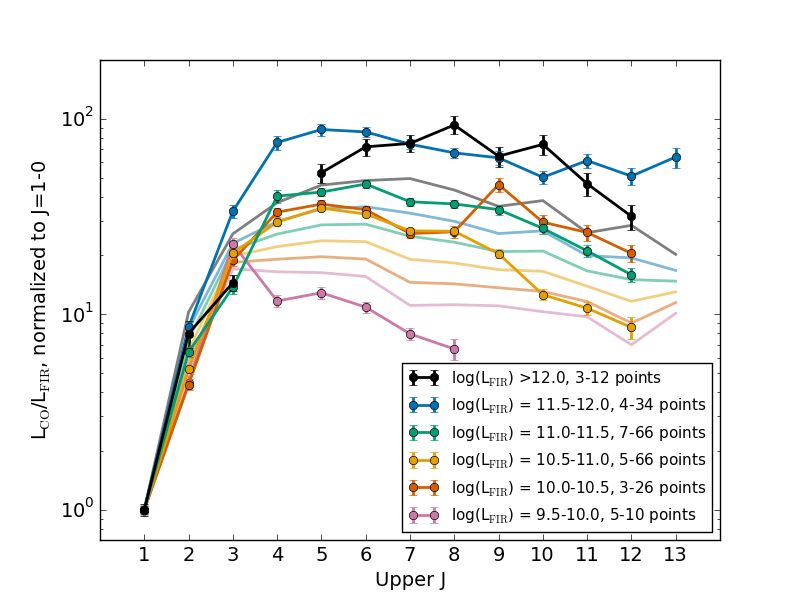}
\includegraphics[width=\columnwidth]{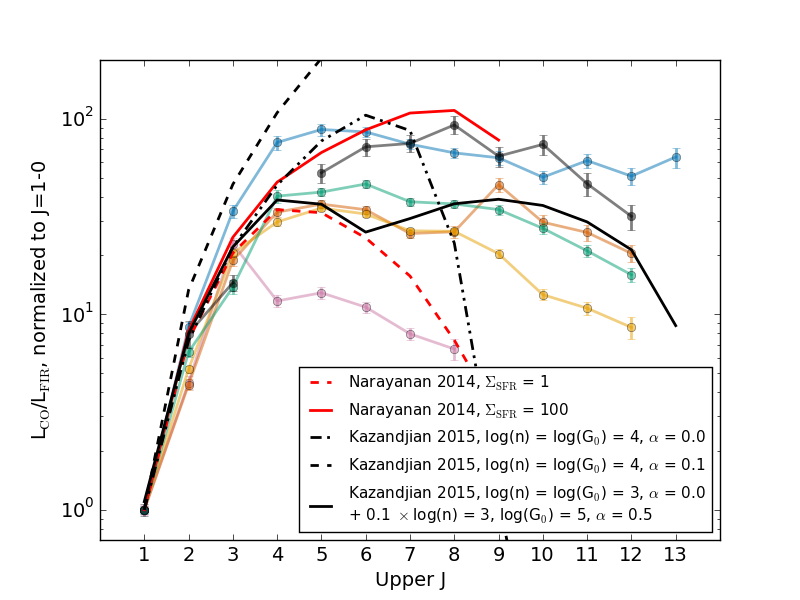} 
\caption[Average SLEDs by \lfir\ Ranges]{Top: average SLEDs by \lfir\ ranges. For \lfir\ bin ranges shown in the legend, the value of \lco/\lfir\ was averaged if measurements existed for at least 3 galaxies. All SLEDs were then divided by the value of the \lco (\jone)/\lfir\ line to demonstrate the difference in relative excitation (shape) of the CO ladder. The number of data points used in each SLED may change with each line, which is why a range is given in the legend. The highest luminosity bin is dominated by more distant galaxies, where the CO \jfour\ line is likely to be redshifted out of the FTS band, which is why that black data point is missing. The lighter lines 
with no data markers indicate the predictions from the slopes in Table \ref{tbl:slopes} for the center of each log bin. 
Bottom: comparison to theoretical models. The red model predictions are from \citet{Narayanan2014}, for SFR surface densities of 1 (dashed) and 100 (solid) \ms\ kpc$^{-2}$ yr$^{-1}$ (up to \jnine). The black model predictions are from \citet{Kazandjian2015} for solar metallicity and $A_V = 10$.  The first two predictions are for $n_{\rm gas} = 10^4$, $G_0 = 10^4$, $\alpha = 0$ (dash-dot-dot, which drops quickly) and $\alpha = 0.1$ (dashed, which rises too high).  The solid black line the sum of two models, $n_{\rm gas} = 10^3$, $G_0 = 10^3$, $\alpha = 0$ to fit the lowest-$J$ lines, and then the model of $n_{\rm gas} = 10^3$, $G_0 = 10^5$, $\alpha = 0.5$, multiplied by 0.1, to attempt to (though not well) reproduce a flat mid- to high-$J$ spectrum at less than 100. The lines with data points correspond to the average SLEDs in the top panel.
Note: \lco $\propto \nu^3$\lprimeco, \lco $\propto \nu I_{CO}$ [Jy \kms].
\label{fig:averagesleds}}
\end{figure}

Within galaxy-wide log(\lfir) ranges of approximately 0.5 dex, we compiled weighted-averaged (\lco/\lfir)/(\lco(\jone)/\lfir) ratios, 
presented in Figure \ref{fig:averagesleds} (top; see caption for note about conversion to \lprimeco). The SLEDs are normalized to the \jone\ line to show the relative excitation across the CO ladder. The CO \jone\ line measures emission from the same type of cold, ubiquitous molecular gas found throughout many types of galaxies.  There are four main trends to examine in this plot: First, there is a trend with increasing \lfir\ towards much more high-$J$ CO luminosity compared to \jone. This indicates a greater energy input relative to typical PDRs, to explain the high-$J$ emission. Second, the location of the CO luminosity peak moves to higher-$J$ with higher \lfir. 
Third, the slope of the mid- to high-$J$ CO emission SLED becomes flatter with increasing \lfir. However, all the SLED slopes are relatively flat, none show an extreme drop-off. 
Fourth, the values of mid- to high-$J$ CO relative to CO \jone\ in luminosity only range from about 10 to 100 across all lines.\footnote{
In brightness temperature units (\lprimeco), this is equivalent to 0.16-1.6 for \jfour\ and 0.0046-0.046 for \jthirteen, because \lco $\propto \nu^3$ \lprimeco.}
This is consistent with the CO SLED compilation shown in Figure 5 of K14, which showed high-$J$/\jone\ ratios from about 5 to 100 (with one outlier, NGC6240, at 250 at its peak, see \citet{Meijerink2013}). These last three trends are indicative of a higher {\it average} pressure (product of kinetic temperature and density) required to explain the shape of the CO SLED. We emphasize {\it average} because all of the CO emission is the sum of a gradient of conditions from different environments.

Figure \ref{fig:averagesleds} (top) also shows a difference between the bin-averaged SLEDs (with data points) and the SLEDs that would be predicted simply from our log(\lfir) vs. log(\lprimeco) slopes (light colored, no data points). The predicted SLEDs span a narrower range than the bin-averaged SLEDs. Describing each CO line by a single relationship over the entire sample averages over the very real differences in populations, such as those described in Section \ref{sec:analysis:slopes} (e.g. different \lco/\lfir\ slopes at higher luminosities). Also, the actual weighted averages are often influenced by high S/N galaxies that may not represent the whole bin, but still illustrate the overall trend. Either way of examining the data, which we do more in the following section, shows a relatively narrow range of high-$J$ to CO \jone\ ratios compared to expectations from some theoretical models.


\begin{figure*}  
\centering
\includegraphics[width=7in]{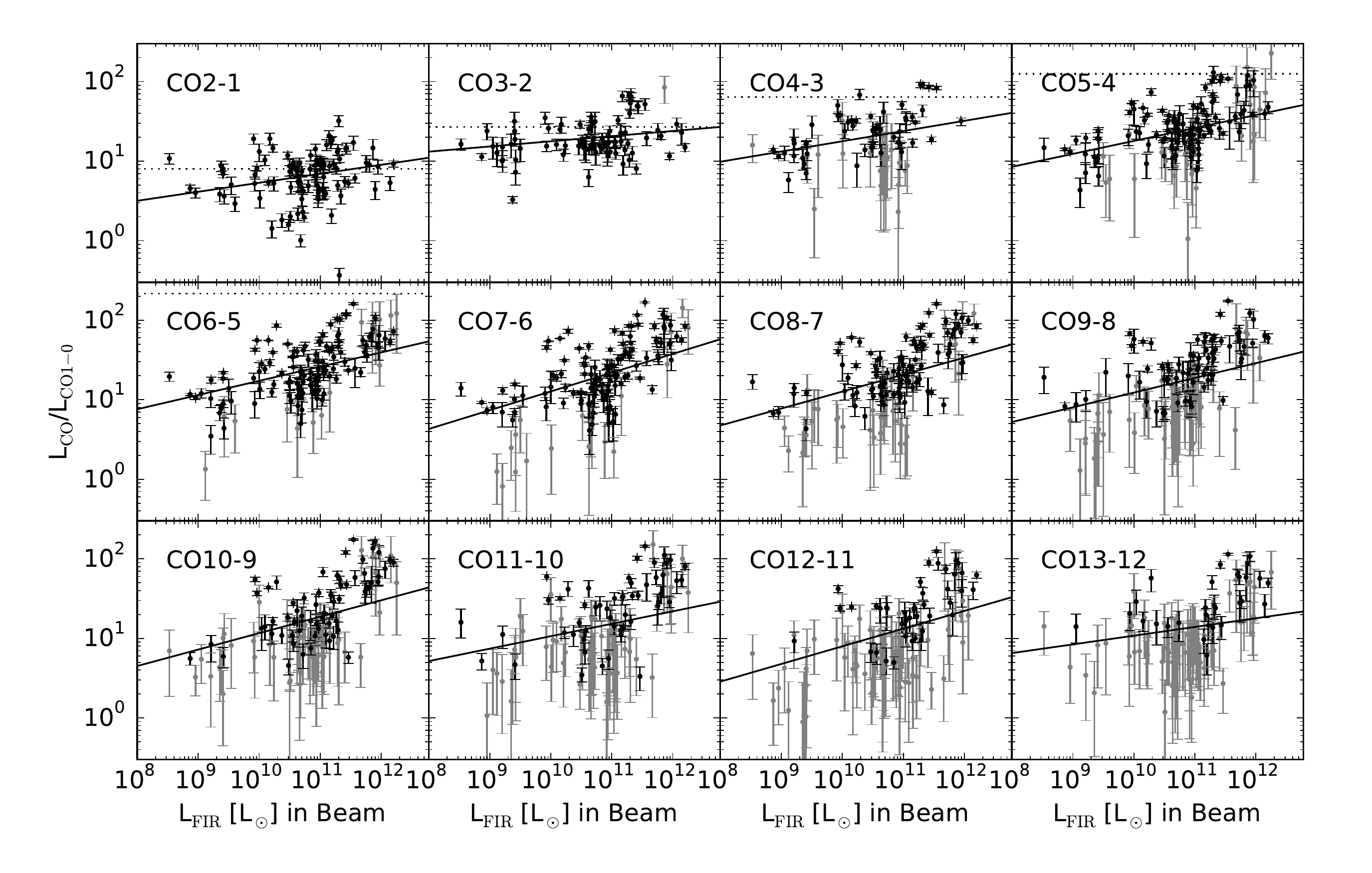} 
\caption[$L_{\rm CO}/L_{\rm CO,1-0}$ vs. \lfir]{$L_{\rm CO}/L_{\rm CO,1-0}$ vs. \lfir. The x-axis is the \lfir\ in the beam for comparison to the CO measurement (see Section \ref{sec:obs:sourcebeam}. The y-axis is the luminosity ratio compared to \jone\ for each line. Black data points indicate $3\sigma$ detections in both lines; gray indicate less than 3 sigma in the higher-$J$ line.  The dotted line denotes $J_{\rm upper}^3$, the theoretical level for thermalized emission (off the top of the y-axis after \jsix).  The solid line denotes the ratios based on the line fits in Table \ref{tbl:slopes} (not a fit to these data, and not utilizing the same data; this plot relies on 1:1 matching of CO lines with a \jone\ measurement from the same galaxy, and multiple low-$J$ lines from the literature are averaged together).  Line ratios across the range of \lfir\ remain generally the same across lines above \jsix, consistently between 1 or 10 to 100.
\label{fig:lineratios}}
\end{figure*}

\section{Discussion}\label{sec:disc}

In the previous section, we explained the observational and theoretical motivations for fitting the slopes of log(\lfir) vs. log(\lprimeco). We now place the trends uncovered in the previous section in the context of theoretical predictions for CO emission. 
One may initially attempt to explain our linear slopes for the mid- to high-$J$ CO lines using the same logic as already presented in Section \ref{sec:analysis} to describe the linear slope of the HCN \jone\ line (because the critical densities increase with upper-$J$ level).
This cannot naively be touted as the explanation for our linear slopes because the assumption of isothermality is not correct \citep[Section 4.3.2]{Krumholz2007}. Their models rely on the gas temperature being lower than the upper-state energy level of the transition. The modeling of K14 has already shown that the kinetic temperature of the high-$J$ emitting gas is quite high and that the lines are not thermalized; {\it both} temperature and density play an important role in the emission. The conditions are not uniform, and this sensitivity to both temperature and density (and invalidity of LTE) is why the high-$J$ CO emission is not linearly related to warm molecular mass.  Such a relationship (between high-$J$ CO emission and mass) was not found in K14 for those reasons. Excitation modeling assuming statistical equilibrium parameterized by pressure (temperature $\times$ density) can better illuminate the physical conditions of the molecular gas.

\citet{Narayanan2008b} extended the argument of \citet{Krumholz2007} by applying excitation and radiative transfer calculations to hydrodynamical simulations of disk and merger galaxies. They matched the known relations and predicted those for the (at the time) unobserved mid-$J$ CO to be less than linear, decreasing from about 0.6 to 0.4 from \jfour\ to \jseven, based on the density and temperature distributions derived from their simulations.  For these lines, the decrease in the slope is due to the fraction of the emission dominated by subthermal excitation increasing with critical density of the tracer. This reiterates the above point that any one individual high-$J$ CO line is a poor tracer of mass. We do not find such low slopes, and also do not discern any trend after \jfive. In these relations, they strictly considered the SFR, not \lfir\ (which we and others use as a proxy). 

\citet{Narayanan2014} showed that the CO SLED shape in their models can be parameterized by star formation rate surface density. Their simulations only considered heating by far ultraviolet (FUV) photons, cosmic rays, and energy exchange with dust. FUV heating is the driving force behind photodissociation regions (PDRs). Two examples are shown in Figure \ref{fig:averagesleds} (bottom); some star formation rate surface densities are a qualitative match to our highest luminosity (and therefore highest SFR) galaxies, but only up through the mid-$J$ lines. Though they do not predict above \jeight, the models imply a sharp drop-off would begin above this line. However, even in galaxies of two orders of magnitude lower luminosity, we still find quite flat SLED slopes in the high-$J$ lines that are not well matched by the models, which drop off steeply after mid-$J$.

As already mentioned in the introduction, typical PDRs cannot account for the highly excited CO SLEDs seen in many {\it Herschel} spectra. 
The need for energy sources beyond FUV photons was demonstrated with data available from the ground \citep{Papadopoulos2012}, but {\it Herschel} has made it even more clear in a number of individual studies and surveys \citep[][and individual references within all three]{Kamenetzky2014, Lu2014,Pereira-Santaella2014}. 
 \citet{Lu2014} divided the shock scenario into two categories: those associated with current star formation (supernovae, winds) and those associated with other non-SF-related phenomena (AGN-driven outflows, radio jets, or galaxy-galaxy collisions). They found SF-dominated galaxies all had similar ratios of total mid-J (\jfive, \jsix, \jseven, \jeight, and \jten) CO luminosity to \lir, whereas galaxies with non-SF-related shocks and high AGN contribution had higher and lower ratios, respectively. 

\citet{Kazandjian2015} extended the treatment of PDR models to include varying degrees of influence from mechanical heating and predicted the CO emission. They found that high-$J$ CO line ratios are especially sensitive to mechanical heating; the same cloud conditions (parameterized as the gas density and FUV irradiating flux, $n$ and $G_0$) can produce dramatically different CO SLEDs with only a few percent of the total heating contributed by mechanical energy ($\alpha$ = $\Gamma_{\rm mechanical}$/$\Gamma_{\rm total}$) such as turbulence and winds (see their Figure 6). 
We examined the CO high-$J$/\jone\ luminosity ratios and 
attempted to compare the \citet{Kazandjian2015} models to our overall trends. While the addition of mechanical energy ($\alpha > 0$) can result in flatter high-$J$ spectra, we find these ratios dramatically overpredict the high-$J$ luminosity relative to the \jone\ line. Two examples (out of a much larger parameter space) are shown in Figure \ref{fig:averagesleds} (bottom) to illustrate the impact of $\alpha$. With no mechanical heating, the shape of the SLED falls down too dramatically at mid-$J$. With the addition of mechanical heating, though the shape is flatter, it rises far above the ratios we find (and off the top of the plot), which generally do not go above 100 $L_{\rm CO,1-0}$, and never above 180 $L_{\rm CO,1-0}$ for 3$\sigma$ detections.
We can somewhat reproduce the average shape by combining multiple models with the higher excitation component reduced by a large percentage, indicating a negligible amount of CO \jone\ emission from this component. One example is shown in Figure \ref{fig:averagesleds}, but more detailed comparisons to the models examining all the possible parameter space of the models is required.

Examination of Figure \ref{fig:averagesleds} indicates that our trends derived from the \lfir/\lprimeco\ fitting (solid, light colored lines) may not be representative of the population as a whole, given the variance in the average SLEDs from these trends. We therefore also examined the line ratios $L_{\rm CO}/L_{\rm CO,1-0}$ as a function of \lfir\ in Figure \ref{fig:lineratios}. These data are different than that fit in Figure \ref{fig:slopes} because a) it relies on the high-$J$ lines for a given spectrum having a corresponding CO \jone\ line and b) it averages the CO \jone\ luminosity when multiple measurements are available (though all are referenced to the same beam size). 
The same behavior we see in the average SLEDs is still present; the luminosity of the high-$J$ lines relative to CO \jone\ only varies by about 1.7 orders of magnitude across the range of \lfir. This is true for each of the CO lines, which is why the slopes in the average SLEDs are all relatively flat.
This way of looking at the data also illustrates the differences in population, e.g. (U)LIRGs lying above the average trendline, which are averaged out when fitting the log(\lfir)/log(\lprimeco) slopes as in Figure \ref{fig:slopes}.

We used the same method as in K14 to conduct 2-component non-LTE likelihood modeling of the average CO SLEDs by the same \lfir\ bins as in Figure \ref{fig:averagesleds}. We describe the SLED as a sum of two components of gas, each described with 4 parameters: kinetic temperature, density of molecular hydrogen, column density of CO, and area filling factor. While the molecular gas exists in a continuum of conditions, two components is the simplest description of these conditions (see further discussion in K14). The product of temperature and density, the pressure P/k in K cm$^{-3}$, largely determines the relative shape of the SLED. The product of column density and area is proportional to the mass, which determines the total emission (and as previously discussed, in the case of the cold gas only, is often considered directly proportional to CO \jone). 

The trends with \lfir\ for the predictions from the \lco/\lfir\ slopes (light solid colors in Figure \ref{fig:averagesleds}) can be mostly foreseen from the shapes alone; the pressure, mass, and luminosity for both the warm and cold components increase slightly with increasing \lfir, but overall the physical conditions are not too dissimilar up through log(\lfir) = 12. The ratio of warm to cold CO luminosity is about 45-65 for all bins log(\lfir) $<$ 12, but higher for ULIRGs, around 200. 
The warm/cold component pressure on average drops from about 60 to 50 from the log(\lfir) = 9.5-10 bin to the log(\lfir) = 11.5-12.0 bin, but is only about 25 for the highest bin of log(\lfir) $>$ 12.
The cold component pressures range from log(P/k [K cm${-3}$]) = 4 to 4.5, but 4.7 for ULIRGs, and the warm component from 5.7 - 6. Within this parameter, there is substantial degeneracy between temperature and density.  
Finally, the warm/cold mass ratio is about 0.2, but only 0.1 for the ULIRG bin. 
This means that while both the cold and warm components have higher pressure overall, more of the mid-$J$ emission in the ULIRGs can be explained by higher bulk excitation of the total molecular gas (most of which is cold). This could have implications for the stellar initial mass function (IMF) and its subsequent effect on the surrounding gas.

In summary, we find linear slopes between mid- to high-$J$ log(\lprimeco) and log(\lfir). Because this CO emission is not thermalized, it should not be used as a proxy for molecular mass or a strict K-S relation. Such slopes are inconsistent with the hydrodynamical simulations of \citet{Narayanan2008b}. The relative luminosity of high-$J$ CO to \jone\ slightly correlates with \lfir, but only ranges from about 10 to 100 for \jsix\ through \jthirteen\ when a single power law describes each line. When examining the full sample (Figure \ref{fig:lineratios}), this range varies from a few to 180.  Across the range of \lfir\ here, the SLEDs relative to CO \jone\ are fairly flat across these lines. Neither hydrodynamical nor PDR+mechanical heating models reproduce this trend when used as a single descriptor of the galaxy-wide emission. Combinations of such descriptors, which essentially adjusts the relative contributions of molecular gas components, could better describe the SLEDs.


\section{Conclusions}\label{sec:conclusions}

We have presented a catalog of all CO, \ci, and \nii\ lines available from extragalactic spectra from the {\it Herschel} SPIRE Fourier Transform Spectrometer.

\begin{enumerate}
\item Our catalog includes a uniform treatment for source/beam coupling correction and proper estimates of the probability distribution functions (PDFs) of line flux measurements given the highly correlated nature of the Fourier Transform Spectrometer.
\item Relations of the form log(\lfir) = a log(\lprimeco) + b show linear slopes over multiple orders of magnitude for mid- to high-$J$ CO lines, and slightly sublinear slopes if restricted to (U)LIRGs.
\item Average SLEDs show increasing mid- to high-$J$ CO luminosity relative to CO \jone, from a few to $\sim$ 100, with increasing \lfir. Even for the most luminous local galaxies, the high-$J$ to \jone\ ratios do not exceed 180.
\item The luminosities relative to CO \jone\ remain relatively flat from \jsix\ through \jthirteen, across many orders of magnitude of \lfir. 
\item Current theoretical models do not reproduce such flat SLEDs with ratios $<$ 180 across such a large range of galaxy luminosity.
\item Preliminary RADEX modeling shows that more of the mid-$J$ emission in ULIRGs can be attributed to higher bulk excitation of the total molecular gas, not just isolated emission from high excitation gas.
\end{enumerate}

Future work will include the detailed, two-component excitation modeling of galaxy spectra with at least eight of the thirteen CO lines shown here, as in K14. 

\acknowledgments

We thank the anonymous referee for a thorough and helpful report.
This material is based upon work supported by the National Science Foundation under Grant Number AST-1402193 and 
by NASA under award number NNX13AL16G. 
We utilized multiple publicly available software packages in addition to those already credited in the text, such as astropy, astroquery, pyspeckit, and 
Dave Green's ``cubehelix" colormap.
We acknowledge the usage of the HyperLeda database (http://leda.univ-lyon1.fr).
SPIRE has been developed by a consortium of institutes led by Cardiff University (UK) and including Univ. Lethbridge (Canada); 
NAOC (China); CEA, LAM (France); IFSI, Univ. Padua (Italy); IAC (Spain); Stockholm Observatory (Sweden); Imperial College London, 
RAL, UCL-MSSL, UKATC, Univ. Sussex (UK); and Caltech, JPL, NHSC, Univ. Colorado (USA). This development has been supported by national 
funding agencies: CSA (Canada); NAOC (China); CEA, CNES, CNRS (France); ASI (Italy); MCINN (Spain); SNSB (Sweden); STFC (UK); and NASA (USA).
We would like to thank Rosalind Hopwood for useful guidance with HIPE reprocessing and Karin Sandstrom for sharing planetary calibration observations.

{\it Facilities:} \facility{Herschel (SPIRE)}, \facility{ARO:SMT}, \facility{ARO:12m}

\appendix

\section{Illustrated Example of Line Fitting Procedure}\label{sec:linefitexample}

We chose NGC4388 as an example of a semi-extended galaxy with a fairly good spectrum that degrades in signal/noise by the time it reaches the higher-$J$ CO lines. 

\begin{figure*}  
\centering
\includegraphics[width=7in]{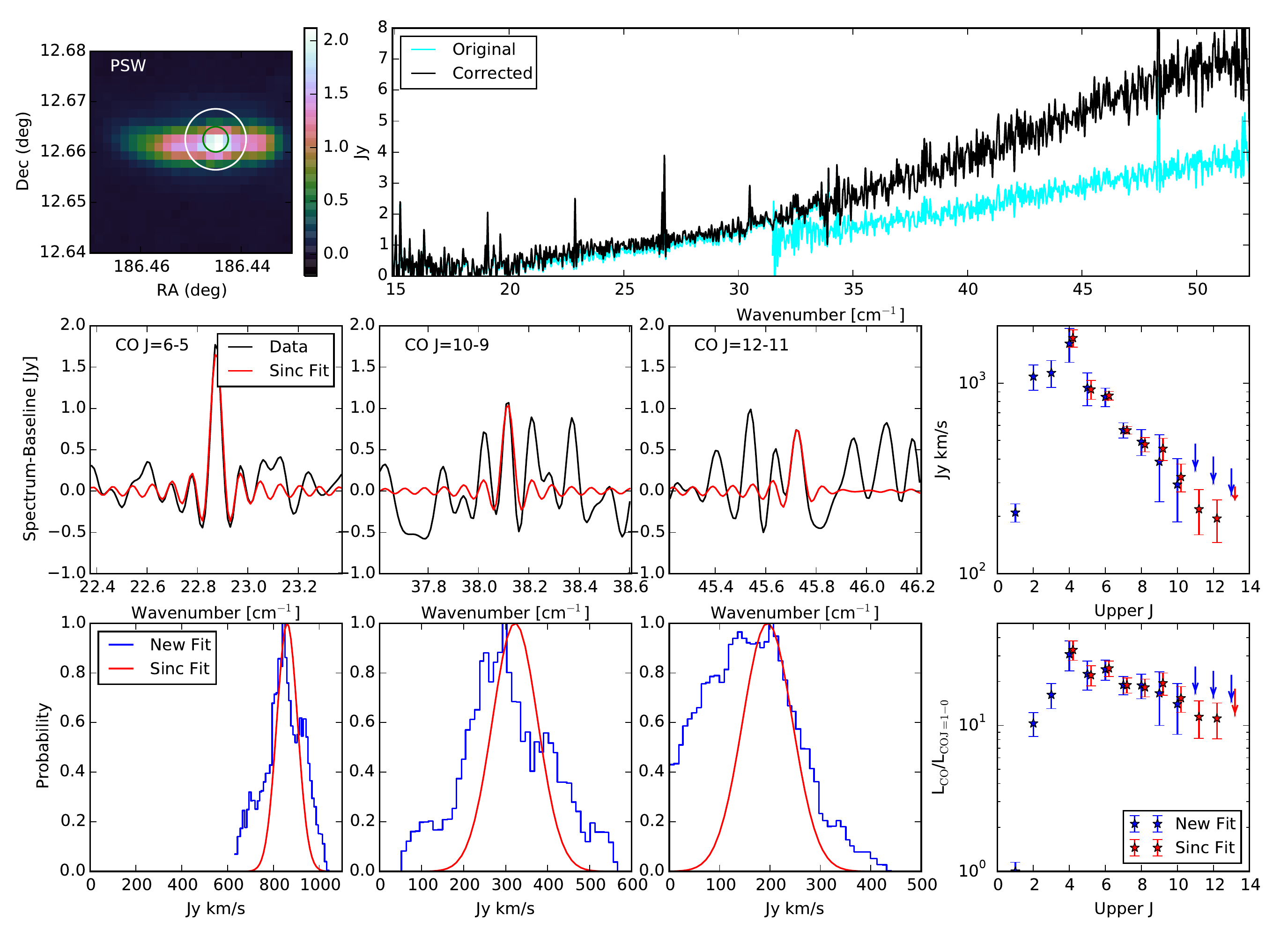} 
\caption[Line Fitting of NGC4388]{Line Fitting of NGC4388.
{\it Top Left:} The PSW map of NGC4388 illustrates that this galaxy is extended relative to the largest and smallest spectrometer FWHM (white, green circle). As a result, the original spectrum (cyan in {\it Top Right}) shows a noticeable gap between the SLW and SSW bands. The source/beam size and photometer-matched corrected spectrum is shown in black (described in Section \ref{sec:obs:sourcebeam}).
{\it Middle Row, Left Three Columns:} Zoomed-in views of the baseline-subtracted spectra (black) for three lines, and the best-fit line profile using the FTFitter (red).
{\it Bottom Row, Left Three Columns:} Probability distribution functions of the integrated line fits of the row above, for the FTFitter (red, assuming a Gaussian profile and using the fit and parameter error as median and sigma), and for our new method (blue). For some lines, the distribution function is much wider, and/or more tending towards zero than the least squares fitting routine would reveal, given the surrounding noise profile. 
{\it Right Column:} Resulting SLEDs for the line fits in Jy km/s ({\it Middle Row}) and in \lco/$L_{\rm CO,1-0}$ ({\it Bottom Row}). The original fits are shown in red, the SLED we use for fitting from our new method is in blue.
\label{fig:galaxyexample}}
\end{figure*}

As an overview, the top rows of Figure \ref{fig:galaxyexample} illustrate the source/beam correction described in Section \ref{sec:obs:sourcebeam}. 
The photometer PSW map shows that the emission is somewhat extended relative to the SPIRE FTS beam sizes, which results in a discontinuity in the original spectrum (cyan, right plot). The corrected spectrum removes this discontinuity, and shows the flux emitted in a 43\farcs5 beam.
The bottom two rows illustrate high signal/noise (first column) vs. poor signal/noise (next two columns) CO lines. Fitting these two lines with a least-squares fitting routine, such as the FTFitter, often produce integrated fluxes of signal/noise greater than 3, because the ``ringing" in the spectrum is well-fit by the intrinsic line profile of the spectrometer. However, inspection of the lines should lead one to question why the surrounding ripple peaks are not also high detections of other lines; none of which are expected to be nearly the intensity of CO. The resulting probability distribution functions in the bottom rows are thus wider and more heavily weighted towards zero.

To describe how the blue PDFs in the aforementioned figure were created, we focus on the CO \jtwelve\ line, for which the FTFitter returns an integrated flux of 0.0298 $\pm$  0.0075 Jy cm$^{-1}$ centered at 
45.73 cm$^{-1}$. For this procedure, we consider the frequency range $\pm$ 2 cm $^{-1}$ from this center, masking out $\pm$ one line profile FWHM (0.048 cm$^{-1}$) around any CO, \ci, and \nii\ lines in this region 
(in this case, only the CO \jtwelve\ line itself). We create a grid of injected line amplitudes, $f_{true}$ from 0 to 0.067 (the range is defined by the minimum of 0 or the flux - 5$\sigma$ to the flux + 5$\sigma$).
For each amplitude, we inject a sinc function of that amplitude at a location within our frequency range and refit the spectrum, recording the total measured integrated flux. This procedure is done at evenly sampled frequencies, every 0.048 cm$^{-1}$, over the frequency range (about 80 samples if no other lines are present nearby). For this input amplitude, we now have a histogram of measured amplitudes, $f_{obs}$. All together, we now have a two-dimensional map of input fluxes vs. measured fluxes $P(f_{obs} | f_{true})$, from which we can back out the probability of the input flux given our measured flux. The blue PDF shown in Figure \ref{fig:galaxyexample} is a slice of this map at measured flux of 0.0298 Jy cm$^{-1}$ ($P(f_{true}|0.0298$); in other words, it is the distribution of input fluxes that produced a measured flux of 0.0298 Jy cm$^{-1}$ in this frequency range.

\bibliography{/Users/julia/papers/bibtex/COFluxes,/Users/julia/papers/bibtex/Dust,/Users/julia/papers/bibtex/Herschel,/Users/julia/papers/bibtex/M82,/Users/julia/papers/bibtex/Molecular,/Users/julia/papers/bibtex/Z-Spec,/Users/julia/papers/bibtex/Methods}

\end{document}